\documentclass[twocolumn]{autart}    % Enable this line and disable the 
\usepackage{amsmath,amssymb,amsfonts}
\usepackage{algorithm,algorithmic}
\usepackage{graphicx}
\usepackage{textcomp}
\usepackage{url}
\usepackage{color}

\usepackage{lipsum}
\usepackage{epstopdf}

\usepackage{amsopn}
\DeclareMathOperator{\diag}{diag}

\usepackage{enumerate}
\usepackage{cancel}
\usepackage{mathrsfs}
\usepackage{mathdots}
\usepackage{euscript}
\usepackage{amscd}
\usepackage{placeins}
\usepackage{natbib}

\usepackage{subcaption}

\usepackage{tikz}
\usetikzlibrary{arrows,shapes,chains}

\newtheorem{theorem}{Theorem}

\newtheorem{lemma}{Lemma}
\newtheorem{proposition}{Proposition}

\theoremstyle{definition}

\newtheorem{example}{Example}

\theoremstyle{remark}
\newtheorem{remark}{Remark}

\newcommand{\bmat}{\left[ \begin{matrix}}
	\newcommand{\emat}{\end{matrix} \right]}
\newcommand{\innerprod}[2]{\langle{#1},\,{#2}\rangle}

\DeclareMathOperator{\trace}{tr}

\DeclareMathOperator{\relint}{relint}

\DeclareMathOperator{\E}{{\mathbb E}}
\newcommand{\Rbb}{\mathbb R}

\newcommand{\Cbb}{\mathbb C}
\newcommand{\Dbb}{\mathbb D}

\newcommand{\Zbb}{\mathbb Z}

\newcommand{\Tbb}{\mathbb T}

\newcommand{\xb}{\mathbf  x}
\newcommand{\yb}{\mathbf  y}
\newcommand{\sbf}{\mathbf  s}  %\sb already defined
\newcommand{\zb}{\mathbf  z}
\newcommand{\wb}{\mathbf  w}

\newcommand{\hb}{\mathbf  h}

\newcommand{\bb}{\mathbf  b}

\newcommand{\ub}{\mathbf  u}
  
\newcommand{\cb}{\mathbf c}
\newcommand{\qb}{\mathbf q}

\newcommand{\zerob}{\mathbf 0}

\newcommand{\Gb}{\mathbf G}

\newcommand{\Rb}{\mathbf R}

\newcommand{\lambdab}{\boldsymbol{\lambda}}
\newcommand{\rhob}{\boldsymbol{\rho}}
\newcommand{\thetab}{\boldsymbol{\theta}}

\newcommand{\nub}{\boldsymbol{\nu}}
\newcommand{\psib}{\boldsymbol{\psi}}

\DeclareMathOperator{\range}{Range}
\DeclareMathOperator{\rank}{rank}
\newcommand{\Acal}{\mathcal{A}}

\newcommand{\Ical}{\mathcal{I}}
\newcommand{\Dcal}{\mathcal{D}}
\newcommand{\Hcal}{\mathcal{H}}

\renewcommand{\d}{\mathrm{d}}
\renewcommand{\Re}{\mathrm{Re}}

\newcommand{\SNR}{\mathrm{SNR}}

\newcommand{\srm}{\mathrm{s}}
\newcommand{\adj}{\mathrm{adj}}

\newcommand{\pfend}{{\hfill $\Box$}}

\newcommand{\floor}[1]{\lfloor #1 \rfloor}

\newcommand{\RMSE}{\mathrm{RMSE}}
\begin{document}

\begin{frontmatter}

\title{Line Spectral Analysis Using the G-Filter: An Atomic Norm Minimization Approach\thanksref{footnoteinfo}} % Title, preferably not more 

\thanks[footnoteinfo]{This work was supported in part by Shenzhen Science and Technology Program (Grant No.~202206193000001-20220817184157001), and the “Hundred-Talent Program” of Sun Yat-sen University. Corresponding author B.~Zhu. Tel. +86 14748797525. Fax +86(xx) xxxxxxxx.}

\author[SYSU]{Bin Zhu}\ead{zhub26@mail.sysu.edu.cn},   % Add the 
\author[SYSU]{Jiale Tang}\ead{tangjle@mail2.sysu.edu.cn}              % e-mail address 

\address[SYSU]{School of Intelligent Systems Engineering, Sun Yat-sen University, Gongchang Road 66, 518107 Shenzhen, China}

\begin{keyword}                           % Five to ten keywords,  
Line spectral analysis, frequency estimation, G-filter, Carath\'{e}odory--Fej\'{e}r-type decomposition, atomic norm minimization, convex optimization.         % chosen from the IFAC 
\end{keyword}                             % keyword list or with the 
                                          % help of the Automatica 
                                          % keyword wizard

\begin{abstract}                          % Abstract of not more than 200 words.

The area of spectral analysis has a traditional dichotomy between continuous spectra (spectral densities) which correspond to purely nondeterministic processes, and line spectra (Dirac impulses) which represent sinusoids. While the former case is important in the identification of discrete-time linear stochastic systems, the latter case is essential for the analysis and modeling of time series with notable applications in radar systems. In this paper, we develop a novel approach for line spectral estimation which combines ideas of Georgiou's filter banks (G-filters) and atomic norm minimization (ANM), a mainstream method for line spectral analysis in the last decade following the theory of compressed sensing. Such a combination is only possible because a Carath\'{e}odory--Fej\'{e}r-type decomposition is available for the covariance matrix of the filter output. The ANM problem can be characterized via semidefinite programming which can be solved efficiently. As a consequence, our optimization {scheme} can be seen as a substantial generalization of the standard ANM for line spectral estimation. Moreover, our ANM approach with a G-filter has significant advantages over subspace methods because it can work with just one output vector and without \emph{a priori} knowledge about the number of sinusoids in the input. Simulation results show that our approach performs 
{favorably against the standard ANM, the frequency-selective ANM, and standard subspace methods MUSIC and ESPRIT under a variety of parameter configurations}
when the G-filter is suitably designed.
\end{abstract}

\end{frontmatter}

\section{Introduction}\label{sec:intro}

Given a finite number of measurements of a sinusoidal signal, estimating the underlying frequencies
is a classic problem in astrophysics, engineering, and many other scientific fields, having produced a large body of literature, cf.~e.g., \citet{quinn2001estimation,stoica2005spectral}. The frequency-domain formulation of the same problem is known as \emph{line spectral estimation}, since the spectrum of a sinusoidal signal consists of Dirac impulses. The space-domain variant of the problem is also called ``direction-of-arrival estimation'' or ``array processing'' \citep{van2004optimum} which plays a fundamental role in applications such as radars and sonars, and imaging systems \citep{borcea2002imaging}.

In the field of systems and control, the spectral estimation problem is intimately connected to modeling a stationary process {$y(t)$} as the output of a linear time-invariant %(LTI) 
system (called ``shaping filter'' in terms of signal processing) driven by white noise {$w(t)$}, see Fig.~\ref{fig:model_stationary_proc}. The idea goes back to Wiener in the 1940s {which had a major impact on the subsequent development of statistical filtering including the Kalman filter}. However, for a general wide-sense stationary process, such a model is \emph{not} always possible. Indeed, processes $y(t)$ that can be described by Fig.~\ref{fig:model_stationary_proc} are called \emph{purely nondeterministic}, and their spectra are absolutely continuous and log-integrable. In contrast, {sinusoids, even if their phases, amplitudes, and frequencies are random variables, are called \emph{purely deterministic} because} they can be perfectly predicted using the past of the process, cf.~Wold decomposition in e.g., \citet[Ch.~4]{LP15}. It is the latter case that constitutes the focus of the current paper, {that is, we aim to address the spectral estimation problem for sinusoids.}

\begin{figure}[h]
	\centering
	\tikzstyle{int}=[draw, minimum size=2em]
	\tikzstyle{init} = [pin edge={to-,thin,black}]
	\begin{tikzpicture}[node distance=2cm,auto,>=latex']
	\node [int] (a) {$\ W(z)\ $};
	\node (b) [left of=a, coordinate] {};
	\node (c) [right of=a] {};
	%\node [coordinate] (end) [right of=c, node distance=2cm]{};
	\path[->] (b) edge node {$w(t)$} (a);
	\path[->] (a) edge node {$y(t)$} (c);
	%\draw[->] (c) edge node {$p$} (end) ;
	\end{tikzpicture}
	\caption{A {single-input-single-output} linear time-invariant system driven by white noise {$w(t)$}.}
	\label{fig:model_stationary_proc}
\end{figure}

Many approaches have been developed for {the estimation of spectral lines (or equivalently, of frequencies in sinusoids)}, ranging from classic FFT methods which are essentially linear and computationally cheap, to modern subspace methods which use more linear algebraic techniques and enjoy \emph{high-resolution} properties, see \cite{shaghaghi2015subspace,liao2016iterative,di2018space,Picci-Z-2019-FreqEst,Picci-Zhu-SYSID21,Picci-Zhu-2021} for some recent contributions. A third class of methods, which have been extensively investigated in the past decade, are inspired by the literature on ``compressed sensing'' (see e.g., \citet{eldar2012compressed}) and use more sophisticated convex optimization methods. The idea is to view the sinusoids as a \emph{spectrally sparse} signal and use certain norms to promote such sparsity. It turns out that these methods often have better performance when the sample size is small. A representative in this class of methods is known as \emph{atomic norm minimization} (abbreviated as ANM), see the line of research in \cite{tang2013compressed,bhaskar2013atomic,li2015off,yang2015enhancing,yang2016exact,yang2016vandermonde,yang2018fast,yang2018frequency,yang2018sparse,Zhu-M2-LineSpec,yang2024separation}. The ANM method is related to minimization of the total-variation norm for atomic measures \citep{candes2014towards,fernandez2016super}.
It is worth noting that the ANM approach for frequency estimation results in convex semidefinite programs for which there are off-the-shelf solvers. This feature considerably differs from traditional optimization methods based on \emph{Maximum Likelihood} \citep[Sec.~4.3]{stoica2005spectral} which are typically nonconvex and hard to solve, especially when the sample size is small.

The base of subspace methods and compressed sensing methods for frequency estimation is one important mathematical result known as \emph{Carath\'{e}odory--Fej\'{e}r} (abbreviated as C--F) decomposition\footnote{The C--F decomposition is also termed Vandermonde decomposition in the signal processing literature, see e.g., \citet{yang2018sparse}. It appears, however, that such a decomposition has nothing to do with Vandermonde himself other than the fact that the two factors on the left and right are Vandermonde matrices whose columns or rows are geometric progressions.} for positive semidefinite Toeplitz matrices, cf. e.g., \citet{Grenander_Szego, pisarenko1973retrieval}. 
{The result states that any singular positive semidefinite Toeplitz matrix $\Sigma$ of size $n\times n$ admits a \emph{unique} eigen-type decomposition but now the eigenvectors are replaced by vectors of the form $\bmat 1 & e^{i\theta} & \cdots & e^{i(n-1)\theta}\emat^\top$ with $\theta\in [0, 2\pi)$.  If we interpret $\Sigma$ as the signal covariance matrix and $\theta$ as an angular frequency, then the C--F decomposition is naturally compatible with the standard signal model for frequency estimation. Moreover, by encoding the unknown frequencies into $\Sigma$, one can (in a sense) \emph{linearize} the frequency estimation problem by focusing on the estimation of the signal covariance matrix.
In fact, 
this is the main idea behind subspace methods for frequency estimation. While subspace methods extract the signal and noise subspaces in view of the C--F decomposition from the \emph{estimated} covariance matrix $\hat\Sigma$ of the noisy sinusoids, a fundamental difference in compressed sensing methods is that 
the covariance matrix $\Sigma$ is directly used as an optimization variable, which leads to a convex optimization problem. The estimate of the frequencies can then be recovered from the optimal $\Sigma$ in a unique fashion.}

In the papers by \cite{georgiou2000signal,amini2006tunable}, the C--F decomposition has been significantly generalized from {positive semidefinite} Toeplitz matrices to state covariance matrices corresponding to a class of stable linear filter banks which we call ``G-filter'' in this paper. 
Then such a \emph{C--F-type} decomposition has been used to design subspace methods such as MUSIC and ESPRIT for frequency estimation which are compatible with the G-filter structure and show superior performances in comparison with standard subspace methods due to certain \emph{frequency-selective} property of the filter bank. 
This fact motivates the integration of the G-filter and the generalized decomposition result into the ANM framework for frequency estimation so that we can combine the powers from both sides.
On the one hand,  a suitable design of the G-filter can capture and exploit different kinds of band information of the sinusoidal signal, which underpins the improvement of our combined approach over the standard ANM.
On the other hand, the ANM approach usually performs better  than subspace methods with few measurements, and we expect to retain this good property after the incorporation of the G-filter.

The main contributions of this paper are summarized next.
We show that the ANM approach can {incorporate}
an arbitrary G-filter thanks to the C--F-type decomposition, and the resulting optimization {scheme} contains the standard ANM as a special case.
Then the optimization problems are reformulated via semidefinite programming which can be solved in polynomial time.
Conditions of exact frequency localization are given in terms of the dual optimal solution.
Moreover, our ANM approach features a much better small-sample performance in comparison with the G-filter version of {subspace methods}
in \cite{georgiou2000signal,amini2006tunable}.  
In fact, the G-filter version of the ANM approach for frequency estimation requires as few as \emph{one} filtered output to work, as expected in the theoretical development and observed in numerical experiments, while any subspace method would fail in that case simply because it is impossible to estimate the state covariance matrix to satisfactory accuracy.
In addition, extensive simulation studies indicate that the G-filter version of the ANM outperforms the standard ANM \citep{bhaskar2013atomic}, the frequency-selective ANM \citep{yang2018frequency}, and standard subspace methods MUSIC and ESPRIT under various parameter configurations.

The rest of this paper is organized as follows. The classic frequency estimation problem is briefly described in Sec.~\ref{sec:prob}. Georgiou's filter banks and related state covariance matrices are reviewed in Secs.~\ref{sec:G-filter} and \ref{sec:struc_cov_mat}. The C--F-type decomposition for state covariance matrices are discussed in Sec.~\ref{sec:C-F-type_decomp}.
The ANM problem with a G-filter integrated for frequency estimation in the noiseless case is treated in Sec.~\ref{sec:ANM_noiseless}, and the dual problem is analyzed in Sec.~\ref{sec:dual}.
Extension to the case of noisy measurements is sketched in Sec.~\ref{sec:noisy}.
Extensive numerical simulations with all the implementation details are given in Sec.~\ref{sec:sims}.
Finally, Sec.~\ref{sec:conclus} draws the conclusions and lists some open questions.

\subsection{Notation}

Boldface lowercase letters like $\xb$ represent vectors while lightface lowercase letters like $y$ denote scalars. Uppercase letters like $A$ are usually reserved for matrices.
Some specific symbols are explained next.
%{\bf Notation}:
\begin{itemize}
	\item $\Zbb$, the set of all integers.
	\item $\Rbb$, the real line.
	\item $\Cbb$, the complex plane.
	\item $\Dbb$, the open unit disc $\{z \in \Cbb : |z|<1\}$.
	\item $\Tbb$, the unit circle $\{z\in \Cbb : |z|=1\}=\partial \Dbb$ where $\partial$ stands for the boundary of a set.
	\item $(\,\cdot\,)^* = \overline{(\,\cdot\,)^\top}$, the complex conjugate transpose of a vector $\xb\in \Cbb^n$ or a matrix $A\in\Cbb^{m\times n}$.
	\item $(\,\cdot\,)^{-*}$ is a shorthand for $[(\,\cdot\,)^{*}]^{-1} = [(\,\cdot\,)^{-1}]^{*}$.
	\item The inner product in $\Cbb^n$ is defined as $\innerprod{\xb}{\yb}:=\yb^* \xb$, and real inner product is $\innerprod{\xb}{\yb}_{\Rbb}:= \Re\{\innerprod{\xb}{\yb}\}$.
	\item $\|\xb\|:=\sqrt{\innerprod{\xb}{\xb}}$ is the Euclidean norm of a vector $\xb\in \Cbb^n$.
	\item $\Ical$, the interval $[0, 2\pi)$ for normalized angular frequencies with the unit ``radian''.
	\item $\Hcal_n$ denotes the set of Hermitian matrices of size $n$ which is a linear space over the reals.
	\item $A\geq 0$ means that the matrix $A$ is positive semidefinite.
	\item The conjugate of a vector-valued rational function $G(z):=(zI-A)^{-1}\bb$ of $z\in\Cbb$ is the rational function $G^*(z):=\bb^*(z^{-1}I-A^*)^{-1}$. Consequently, we have $\left[G(e^{i\theta})\right]^*=G^*(e^{i\theta})$ for any $\theta\in\Ical$.
\end{itemize}

\section{Frequency estimation problem}\label{sec:prob}

The {scalar} signal $y$ under consideration is the noisy measurements of the superposition of some complex sinusoids (cisoids):
\begin{equation}\label{signal_model}
\begin{aligned}
y(t) & = s(t) + w(t) \\
& = \sum_{k=1}^m a_k \, e^{i \theta_k t} + w(t)
\end{aligned}
\end{equation}
where, $t = 0, 1, \dots, L-1$ {is the discrete time variable}, $s$ stands for the signal component, and $w$ the  additive {complex (scalar)} noise component. The signal component is a sum of $m$ cisoids in which each $a_k\in\Cbb$ is an amplitude, and $\theta_k\in\Ical$ is an unknown (but fixed) normalized angular frequency, $k=1, \dots, m$.
{The value of the positive integer $m$ is also unknown \emph{a priori}.}
The measurement model \eqref{signal_model} can also be written in a matrix form
\begin{equation}\label{y_meas_vec}
\yb := \bmat y(0) \\ \vdots \\ y(L-1) \emat = \underbrace{\bmat G_0(e^{i\theta_1}) & \cdots & G_0(e^{i\theta_m}) \emat \bmat a_1 \\ \vdots \\ a_m \emat}_{\sbf} + \wb
\end{equation}
where the column $G_0(e^{i\theta}) := \bmat 1 & e^{i\theta} & \cdots & e^{i(L-1)\theta}\emat^\top$, and $\wb$ is a noise vector {in $\Cbb^L$}.

\emph{Given $L$ noisy measurements $y(t)$ of the cisoidal signal, the problem is to estimate {the integer $m$ and} the unknown frequencies $\{\theta_k\}$}.
It is apparent that once the frequencies $\{\theta_k\}$ are known, estimation of the amplitudes $\{a_k\}$ is a well understood linear problem. For this reason, the problem of frequency estimation is of the main interest.

\section{Review of the G-filter}\label{sec:G-filter}

The G-filter first proposed in \cite{georgiou2000signal} consists of a state/filtering equation:
\begin{equation}\label{filter_bank}
\xb(t) = A\xb(t-1) + \bb y(t), \quad t\in\Zbb,
\end{equation}
where, the state transition matrix $A\in\Cbb^{n\times n}$ is discrete-time (Schur) stable, i.e., its spectral radius $\rho(A)<1$, the vector $\bb\in\Cbb^n$, and $(A, \bb)$ is a \emph{reachable} pair, namely
\begin{equation}
\rank \bmat \bb & A\bb & \cdots & A^{n-1}\bb\emat = n.
\end{equation}
If we use $z^{-1}$ to denote the delay operator $\xb(t) \mapsto \xb(t-1)$, the transfer function of \eqref{filter_bank} is
\begin{equation}\label{trans_func_filter_bank}
G(z) = \bmat g_1(z) \\ \vdots \\ g_n(z) \emat = (I-z^{-1}A)^{-1} \bb = \sum_{k=0}^{\infty} A^k \bb z^{-k},
\end{equation}
where the last power series converges in a domain containing the complement of the open unit disc $\Dbb^\text{c} = \{ z\in\Cbb : |z|\geq 1\}$. In this domain, $G(z)$ is a holomorphic (analytic) function.

\begin{example}[A delay filter bank]\label{ex_delay_filt_bank}
	Take $A$ as the $n\times n$ companion matrix
	\begin{equation}\label{delay_filter}
	A=\bmat 0 & 1 & 0 & \cdots & 0 \\
	0 & 0 & 1 & \cdots & 0 \\
	\vdots & \vdots & \ddots & \ddots & \vdots \\
	0 & 0 & \cdots & 0 & 1 \\
	0 & 0 & \cdots & 0 & 0\emat
	\quad \text{and} \quad
	\bb =\bmat 0 \\ 0 \\ \vdots \\ 0 \\ 1 \emat,
	\end{equation}
	and we have
	\begin{equation}
	g_k(z) = z^{-n+k}, \quad k=1, 2, \dots, n.
	\end{equation}
	If we take the cisoidal signal \eqref{signal_model} as the input, this filter bank just collects the measurements into a vector of size $n$ without additional processing. If $n=L$, 
	the pair $(A, \bb)$ above leads to $G_0(e^{i\theta})$ in \eqref{y_meas_vec} up to a scaling factor $e^{-i(n-1)\theta}$.
\end{example}

\begin{example}[A first-order filter bank]
	Take $A$ to be a diagonal matrix
	\begin{equation}\label{Cauchy_filter}
	A=\bmat p_1 & 0 & 0 & \cdots & 0 \\
	0 & p_2 & 0 & \cdots & 0 \\
	0 & 0 & p_3 & \cdots & 0 \\
	\vdots & \vdots & \vdots & \ddots & \vdots \\
	0 & 0 & 0 & \cdots & p_n \emat
	\quad \text{and} \quad
	\bb =\bmat 1 \\ 1 \\ 1 \\ \vdots \\ 1 \emat.
	\end{equation}
	Now, each
	\begin{equation}
	g_k(z) = \frac{1}{1-p_k z^{-1}}, \quad k=1, 2, \dots, n,
	\end{equation}
	is a standard Cauchy kernel where $p_k\in\Dbb$ represents the pole of the system.
\end{example}

The above examples indicate that the general form {of} the G-filter \eqref{trans_func_filter_bank} encompasses some interesting objects in systems theory and signal processing. Under the reachability condition for $(A, \bb)$, the matrix $A$ can have an arbitrary Jordan structure, see \citet[Sec.~VII-F]{amini2006tunable}. In this sense, the components $\{g_k(z)\}$ of the transfer function $G(z)$ can be regarded as a class of ``generalized Cauchy kernels''. They constitute a \emph{basis} of a subspace of the \emph{conjugate Hardy space} $\bar{H}_2$.

\section{Structure of state covariance matrices}\label{sec:struc_cov_mat}

The state covariance matrix, in view of \eqref{filter_bank}, is defined as $\Sigma:=\E \left[\xb(t)\xb(t)^*\right]$ which is positive semidefinite by construction. Because of the filtering operation, the matrix $\Sigma$ obeys the integral representation 
\begin{equation}\label{state_cov_mat}
\Sigma = \frac{1}{2\pi} \int_{\Ical} G(e^{i\theta})\, \d\mu_y(\theta)\, G^*(e^{i\theta}),
\end{equation}
where $\d\mu_y$ is a nonnegative measure on $\Ical$ which is understood as the power spectrum of the input signal $y$.
Spectral estimation based on the state covariance matrix has been studied intensively since the beginning of this century with pioneering works by \cite{georgiou2000signal,BGL-THREE-00,Georgiou-01,georgiou2002spectral,Georgiou-02,Georgiou-L-03}, subsequent developments in \cite{georgiou2005solution,amini2006tunable,PavonF-06,Georgiou-06,georgiou2007caratheodory,FPR-08,ramponi2009aglobally,RFP-10-wellposedness,FRT-11,FMP-12,ferrante2012maximum,zorzi2012estimation,Z-14,Z-14rat,zorzi2015interpretation,Z-15,GL-17}, and more recent works such as \cite{baggio2018further,Zhu-Baggio-19,zhu2019well}. In these papers, the problem is to infer the power spectrum $\d\mu_y$ of the input from the state covariance matrix $\Sigma$ which is either assumed known or can be estimated to reasonable accuracy. Most of the works above concentrate on \emph{absolutely continuous} measures, i.e., those which admit a \emph{power spectral density}, while \cite{georgiou2000signal,amini2006tunable,georgiou2007caratheodory} consider spectral lines, a special class of \emph{singular} measures\footnote{The concepts of absolute continuity and singularity of measures can be found e.g., in \cite{rudin1987real}.} which correspond to cisoidal signals. In this work, although we do not assume the availability of $\Sigma$ nor its estimate, the algebraic structure of a state covariance matrix determined by the G-filter, which is briefly reviewed next, still plays a fundamental role in our approach for frequency estimation.

Let us define a linear operator 
\begin{equation}
\Gamma: \d\mu \mapsto \frac{1}{2\pi} \int_{\Ical} G(e^{i\theta})\, \d\mu(\theta)\, G^*(e^{i\theta})
\end{equation}
that sends a \emph{signed} measure $\d\mu$ to a Hermitian matrix of size $n$.
Then \eqref{state_cov_mat} implies that $\Sigma\in\range\Gamma$, and the latter is a linear subspace of $\Hcal_n$. The (Banach space) adjoint operator of $\Gamma$ can be identified as
\begin{equation}
	\Gamma^*: X\mapsto G^*(e^{i\theta}) X G(e^{i\theta})
\end{equation}
with $X\in\Hcal_n$. From the definition of the adjoint operator, we have the orthogonality relation 
\begin{equation}
	(\range\Gamma)^\perp = \ker\Gamma^*:=\{X\in\Hcal_n : G^*(e^{i\theta}) X G(e^{i\theta}) \equiv 0\}
\end{equation}
in which the latter is also a subspace of $\Hcal_n$.
In the case of Example~\ref{ex_delay_filt_bank}, $\range\Gamma$ consists of all Hermitian Toeplitz matrices (of size $n$) which are the object of investigation in the classic C--F theorem. At the same time, $\ker\Gamma^*$ contains Hermitian matrices which have all the diagonals (including subdiagonals and superdiagonals) summing up to zero.

The set membership condition $\Sigma\in\range\Gamma$ characterizes the feasibility of spectral estimation problems based on state covariances, including optimization problems to be formulated in later sections of this paper. A number of tests of such feasibility have been found in \cite{georgiou2002spectral,Georgiou-02}, and some of them are summarized below. In particular, $\Sigma\in\range\Gamma$ is equivalent to the rank condition
\begin{equation}
	\rank\bmat \Sigma-A\Sigma A^* & \bb \\ \bb^* & 0\emat 
	= \rank\bmat O & \bb \\ \bb^* & 0\emat =2,
\end{equation}
or the condition that the equality
\begin{equation}\label{charat_range_Gamma}
\Sigma - A\Sigma A^* = \bb \hb^* + \hb \bb^*
\end{equation}
holds for some $\hb\in\Cbb^n$.

\section{Carath\'{e}odory--Fej\'{e}r-type decomposition of state covariance matrices}\label{sec:C-F-type_decomp}

In this section, we {summarize two known results from \cite{georgiou2000signal} into a Carath\'{e}odory--Fej\'{e}r-type decomposition for state covariance matrices $\Sigma$ of the form \eqref{state_cov_mat}, which suits our purpose of theoretical development in subsequent sections.}

\begin{theorem}[{\citet{georgiou2000signal}}]\label{thm:C-F_type_decomp}
%\begin{theorem}[C--F-type decomposition]\label{thm:C-F_type_decomp}
	Let $\Sigma\geq 0$ be a state covariance matrix in the sense of \eqref{state_cov_mat} having rank $r{<} n$. Then it admits a {unique} decomposition of the form
	\begin{equation}\label{C-F-type-decomp}
	\begin{aligned}
	\Sigma & = \sum_{k=1}^{r} \rho_k G(e^{i\theta_k}) G^*(e^{i\theta_k}) \\
	 & = \bmat G(e^{i\theta_1}) & \cdots & G(e^{i\theta_r}) \emat
	 \bmat \rho_1 & & \\ & \ddots & \\ & & \rho_r \\ \emat
	 \bmat G^*(e^{i\theta_1}) \\ \vdots \\ G^*(e^{i\theta_r}) \emat
	\end{aligned}
	\end{equation}
	where {each $\rho_k>0$, and the parameters in $\{\theta_k\}_{k=1}^r\subset\Ical$ are pairwise distinct}.
\end{theorem}

\begin{pf}
	In the case of $r<n$, the existence of a C--F-type decomposition for $\Sigma$ is declared in Proposition 1 of \cite{georgiou2000signal}. 
	The uniqueness follows from Lemma~3 in the same paper which states that {for any positive integer $m\leq n$,} the matrix
	\begin{equation}
	\Gb(\theta_1, \dots, \theta_m) := \bmat G(e^{i\theta_1}) & G(e^{i\theta_2}) & \cdots & G(e^{i\theta_m}) \emat
	\end{equation}
	has linearly independent columns as long as $\{\theta_k\}$ are pairwise distinct.
	Consequently, we have 
	\begin{equation}
	\range\Sigma = \range \Gb(\thetab)  %\bmat G(e^{i\theta_1}) & \cdots & G(e^{i\theta_r}) \emat.
	\end{equation}
	where $\thetab = (\theta_1, \dots, \theta_r)$ {such that $r<n$}.
	Now, suppose that there is another decomposition of the form \eqref{C-F-type-decomp} with parameters $\{\theta'_k, \rho'_k\}_{k=1}^r$.
	Then it must happen that
	\begin{equation}
	\range \Gb(\thetab) = \range \Gb(\thetab').
	\end{equation}
	Due to the linear independence result, the set of frequencies $\{\theta'_k\}$ is necessarily identical to $\{\theta_k\}$, and thus $\{\rho'_k\}$ and $\{\rho_k\}$ are identical as well.  
    \hfill $\Box$
\end{pf}

\begin{remark}
    In the case of a full-rank (i.e., positive definite) $\Sigma$, the C-F-type decomposition  is \emph{not} unique, and a proof can be constructed following the lines in \citet[p.~540]{yang2018sparse}. This point, however, will not be pursued here since in the remaining part of the paper we are only interested in the rank-deficient case.
\end{remark}

In the rank-deficient case, the unique C--F-type decomposition described above is also numerically computable, and this point has been summarized in \citet[Proposition 2]{georgiou2000signal}. More precisely, we first compute the spectral decomposition of $\Sigma$:
\begin{equation}\label{spec_decomp}
\Sigma = U \diag\{\lambda_1, \cdots, \lambda_r, 0, \dots, 0\} U^*
\end{equation}
where $U$ is unitary and $\lambda_k>0$ for $k=1, \dots, r$. Let $\ub_k$ be the $k$-th column of $U$, and partition the eigenvectors as 
\begin{equation}
U = \bmat U_{1:r} & U_{r+1:n}\emat
\end{equation}
where the symbol $U_{k:\ell}$ for $k\leq \ell$ denotes a matrix whose columns are $\ub_k, \ub_{k+1}, \dots, \ub_\ell$.
Then, we construct a rational function which is nonnegative on $\Tbb$:
\begin{equation}\label{rat_func_sym}
\begin{aligned}
d(z, z^{-1}) & = G^*(z) U_{r+1:n}\, U_{r+1:n}^* G(z) \\
 & = \bb^* (I-zA^*)^{-1} U_{r+1:n}\, U_{r+1:n}^* (I-z^{-1}A)^{-1} \bb.
\end{aligned}
\end{equation}
The parameters $\{\theta_k\}$ in \eqref{C-F-type-decomp} correspond to the distinct roots of $d(z, z^{-1})$ on the unit circle, and there are exactly $r$ such roots, i.e., of the form $e^{i\theta_k}$.

Once $\{\theta_k\}$ are obtained, the parameters $\{\rho_k\}$ are determined via linear algebra as follows.
From the spectral decomposition \eqref{spec_decomp} and the C--F-type decomposition \eqref{C-F-type-decomp} of the same matrix $\Sigma$, we can easily form two \emph{rank factorizations} of the form $\Sigma = V_k V_k^*$ with $V_k\in \Cbb^{n\times r}$ for $k=1, 2$. According to \citet[Theorem 7.3.11]{horn2013matrix}, the two factors are related via a unitary matrix $W$, namely $V_1 = V_2 W$ which can be written out as
\begin{equation}
%\begin{aligned}
U_{1:{r}} \diag(\sqrt{\lambdab}) = \Gb(\thetab) \diag(\sqrt{\rhob}) W,
%\end{aligned}
\end{equation}
where, we have introduced the vector $\lambdab=(\lambda_1, \dots, \lambda_r)$ in order to simplify the notation, the square root is taken componentwise, and $\diag(\xb)$ denotes the diagonal matrix whose diagonal elements come from the vector $\xb$.
It then follows that the matrix $\diag(\lambdab^{-{1}/{2}}) T \diag(\rhob^{1/2})$ is unitary, where $T:=U_{1:{r}}^*\Gb(\thetab)$ is invertible because it is the change-of-basis matrix such that $\Gb(\thetab) = U_{1:{r}} T$. After some algebra, we have 
\begin{equation}
\diag(\rhob) = T^{-1} \diag(\lambdab) T^{-*}.
\end{equation}

\begin{example}\label{ex_C-F-type_decomp}%[Illustration of the C--F-type decomposition.]
	This example provides an illustration of the C--F-type decomposition with some computational details. The matrix $\Sigma$ is constructed via \eqref{C-F-type-decomp} in which we take $r=3$, $\thetab=(\theta_1, \theta_2, \theta_3)=(1, 2, 3)$, and $\rhob=(\rho_1, \rho_2, \rho_3) = (8, 4, 2)$. The G-filter has a size $n=20$ and the parameters are determined from a construction in Subsec.~\ref{subsec:construct_G-filter} and used {in simulations of Examples~\ref{ex_separa_freqs}, \ref{ex_high_resol} and \ref{ex_compare_FS}} in Sec.~\ref{sec:sims}. Here we omit these details for the moment and concentrate on the frequency localization property of the symmetric rational function \eqref{rat_func_sym}. 
	Since $U_{r+1:n}\, U_{r+1:n}^*$ in \eqref{rat_func_sym} is a projection matrix, we can scale the rational function to obtain $\bar{d}(e^{i\theta},e^{-i\theta}) := d(e^{i\theta},e^{-i\theta})/\|G(e^{i\theta})\|^2$ which must satisfy the condition $0\leq \bar d(e^{i\theta},e^{-i\theta}) \leq 1$ for all $\theta\in\Ical$. Therefore, the parameters in the vector $\thetab$ are precisely the points of minimum of $\bar{d}$. One can in principle evaluate $\bar{d}$ on a sufficiently dense grid in $\Ical$ and find the $r$ minima using the Matlab function \texttt{mink}. A more stable way to extract the frequencies, however, appears to be the following procedure which consists of three steps:
	\begin{enumerate}
		\item Compute the value of $\bar{d}(e^{i\theta},e^{-i\theta})$ at $N=10^4$ equidistant grid points in $\Ical$, i.e., $\varphi_\ell=\ell\frac{2\pi}{N}, \ell=0, 1, \dots, N-1$. See 
		Fig.~\ref{fig_scale_d_sym_poly_determ_mat} for the graph of the function.
		\item Find those $\varphi_\ell$ such that $\bar{d}(e^{i\varphi_\ell},e^{-i\varphi_\ell})<\varepsilon_1=0.05$.
		\item Run a clustering algorithm\footnote{We simply use \texttt{kmeans} in all the simulations.} on the $\varphi_\ell$ found in Step (2) and the center of each cluster is identified as a frequency in $\thetab$.
	\end{enumerate}

	\begin{figure}
		\centering
		\includegraphics[width=0.35\textwidth]{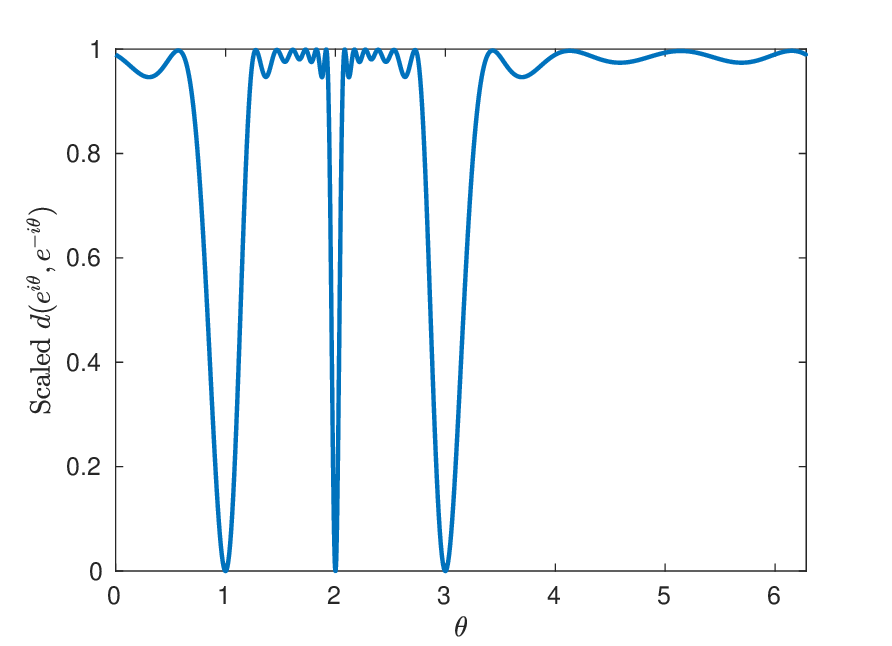}
		\caption{Graph of the scaled rational function $d(e^{i\theta},e^{-i\theta})/\|G(e^{i\theta})\|^2$ which is constructed from a matrix $\Sigma$ of the form \eqref{C-F-type-decomp}.
		The function values are computed on an equidistant grid in $\Ical$ of size $10^4$, and the three minima correspond to the frequency parameters $(\theta_1, \theta_2, \theta_3)=(1, 2, 3)$ of $\Sigma$. }\label{fig_scale_d_sym_poly_determ_mat}
	\end{figure}

	The resulting estimates are
	$\hat\thetab=(0.9987, 1.9999, 3.0012)$, and $\hat\rhob = (8.0152, 4.0000, 2.0034)$ which are close to the true values.
\end{example}

\begin{remark}
	In the Toeplitz case (see Example~\ref{ex_delay_filt_bank}), the computation of the C--F decomposition can be reduced to a \emph{generalized eigenvalue problem} \citep{gurvits2002largest,yang2018sparse,Zhu-M2-LineSpec} without the need to construct the rational function \eqref{rat_func_sym} and to extract its roots. This is only possible due to the special Toeplitz structure. It seems to us that the above procedure must be carried out for the general case dictated by a G-filter.
\end{remark}

\section{Atomic norm minimization approach: the noiseless case}\label{sec:ANM_noiseless}

For the ease of description, let us first assume that the additive noise $w(t)\equiv0$ in \eqref{signal_model}, that is, the measurement $y(t)$ is a precise cisoidal signal. Then $y(t)$ can be represented as the following integral:
\begin{equation}
y(t) = \int_{\Ical} e^{i\theta t} \d\hat{y}(\theta),
\end{equation}
where $\d\hat{y}(\theta) = \sum_{k=1}^m a_k\delta(\theta-\theta_k)\d\theta$ is the \emph{spectral measure} of the signal $y(t)$. Because $\d\hat{y}(\theta)$ is a linear combination of $m$ Dirac measures, it is also called an ``atomic'' measure. \emph{In the remaining part of this paper, we shall assume that the number of spectral lines (cisoids) is less than the size of the G-filter, namely $m<n$}. Notice that this is not a serious restriction because we can always take the filter size $n$ sufficiently large in order to satisfy the strict inequality.

By the expressions of the G-filter \eqref{filter_bank} and \eqref{trans_func_filter_bank}, it is not difficult to derive the relation
\begin{equation}\label{state_atomic_decomp}
\begin{aligned}
\xb(t) = G(z) y(t) & := \int_{\Ical} G(e^{i\theta}) e^{i\theta t} \d\hat{y}(\theta) \\
 & = \sum_{k=1}^m G(e^{i\theta_k}) c_k(t),
\end{aligned}
\end{equation}
where 
\begin{equation}\label{filtered_amp}
	c_k(t) = a_k e^{i\theta_k t}
\end{equation}
are unknown coefficients. In simple terms, the output $\xb$ of the G-filter at time $t$ equals a linear combination of some elements, called ``atoms'' (or dictionary vectors), from the atomic set (also called ``dictionary'')
\begin{equation}\label{dictionary}
\Acal := \{G(e^{i\theta}) : \theta\in \Ical\}.
\end{equation}
Such a linear combination is called an \emph{atomic decomposition}, where the unknown frequency $\theta_k$ is contained in the selected atom $G(e^{i\theta_k})$. Based on this idea, a class of approaches for frequency estimation, inspired by compressed sensing, use the following \emph{atomic norm}:
\begin{equation}\label{atomic_norm}
\begin{aligned}
\|\xb(t)\|_{\Acal} := & \inf_{c_k, \theta_k} \left\{ \sum_{k} |c_k| \|G(e^{i\theta_k})\| : \right. \\
& \left. \xb(t) = \sum_{k} G(e^{i\theta_k}) c_k,\ \theta_k\in \Ical,\ c_k\neq 0 \right\}.
\end{aligned}
\end{equation}
The above object can be interpreted as the spectral version of the (weighted) $\ell_1$ norm. Hence, it is able to promote sparsity in the frequency domain in the sense that the input signal $y(t)$ should contain as few spectral lines (Dirac impulses) as possible. Notice that the dictionary \eqref{dictionary} is obviously over-complete as linear independence is lost among an infinite number of vectors. Such a choice of the dictionary significantly increases the flexibility of atomic norm.

\begin{remark}[Filtering a finite-length input signal]\label{rem_input_filtering}
	Given a cisoidal signal $y(t)$ with $t=0, 1, \dots, L-1$, the filtering operation is just a straightforward implementation of \eqref{filter_bank} with a zero initial condition. The transient effect of the initial condition is remedied by throwing away the first $L_{\srm}$ filtered samples such that $\|A^{L_{\srm}}\|<\varepsilon$ with a predefined threshold $\varepsilon>0$. This is in line with \citet[p.~2664]{amini2006tunable}.
	As a consequence of the above operation, it is rather possible that we can only get \emph{very few}  {``steady-state''} output samples $\xb(t)$, or even \emph{one} sample, such that the subspace methods in \cite{georgiou2000signal,amini2006tunable} do not work.
Indeed, the subspace methods rely on an estimate $\hat\Sigma$ of the covariance matrix $\Sigma=\E [\xb(t)\xb(t)^*]$, and the estimate is typically computed via the averaging scheme
\begin{equation}\label{Sigma_sample_cov}
\hat{\Sigma} = \frac{1}{L_{\xb}} \sum_{t=L_{\srm}}^{L-1} \xb(t) \xb(t)^* 
\end{equation}
with $L_{\xb}=L-L_{\srm}$.
A common assumption of subspace methods is that the covariance matrix estimate $\hat\Sigma$ is nonsingular when the additive noise $w(t)$ in \eqref{signal_model} is (nonzero and) white, or at least the first $m$ eigenvalues of $\hat{\Sigma}$ (ordered by size) are significant enough so that the signal and noise subspaces can be distinguished from each other. However, if the number of available samples is \emph{equal to one}, which is the case considered in this work, the usual covariance matrix estimate in \eqref{Sigma_sample_cov} does not satisfy such an assumption.
\end{remark}

\begin{remark}
	Notice that the number $c_k \|G(e^{i\theta_k})\|$ is the mass of the \emph{normalized} atom $G(e^{i\theta_k}) / \|G(e^{i\theta_k})\|$ in the atomic decomposition in the brace of \eqref{atomic_norm}. In the Toeplitz case (Example~\ref{ex_delay_filt_bank}), $\|G(e^{i\theta})\|=\sqrt{n}$ is a constant. Hence it can be safely removed from the definition of the atomic norm, that is, we can write $\sum_{k} |c_k|$ in place of $\sum_{k} |c_k| \|G(e^{i\theta_k})\|$.
\end{remark}

\emph{Assume that we can only measure the output $\xb(t)$ of the G-filter at a \emph{single} time instance $t$, written simply as $\xb\in\Cbb^n$}. 
How to compute the atomic norm \eqref{atomic_norm} from its definition is not trivial at all. The next result, which can be seen as the G-filter version of \citet[Proposition II.1]{tang2013compressed}, shows that the computation of the atomic norm $\|\xb\|_{\Acal}$ can be converted to a \emph{semidefinite program} (SDP) which can be solved with standard convex optimization algorithms \citep{bv_cvxbook}.

\begin{theorem}\label{thm_AN_noiseless}
	Given one output vector $\xb \equiv \xb(t)$ of the G-filter whose input is the noiseless cisoids, let $p$ be the optimal value of the semidefinite program
	\begin{subequations}\label{AN_semidef_program}
		\begin{align}
		& \underset{\substack{\tau\in\Rbb,\\ \Sigma\in\range\Gamma}}{\text{minimize}}
		& & \frac{1}{2} (\tau + \trace\Sigma) \label{obj_noiseless} \\
		& \text{subject to}
		& & \bmat \tau&\xb^* \\ \xb&\Sigma \emat \geq 0. \label{LMI_constraint}
		\end{align}
	\end{subequations}
	Then the atomic norm $\|\xb\|_\Acal = p$.
\end{theorem}

\begin{pf}
	First we show the inequality $\|\xb\|_\Acal\geq p$. Let $\xb = \sum_{k} G(e^{i\theta_k}) c_k$ be an atomic decomposition of $\xb$. 	
	Define the complex number 
	\begin{equation}\label{u_k_def}
		u_k := c_k \|G(e^{i\theta_k})\| /|c_k|
	\end{equation}
	of modulus $\|G(e^{i\theta_k})\|$ for each $k$, 
	the scalar
	\begin{equation}\label{tau_def}
	\tau:=\sum_{k} |c_k| \|G(e^{i\theta_k})\|,
	\end{equation}  
	and the covariance matrix	
	\begin{equation}\label{Sigma_def}
	\Sigma := \sum_{k} \frac{|c_k|}{\|G(e^{i\theta_k})\|} G(e^{i\theta_k}) G^*(e^{i\theta_k})	
	\end{equation}  
    in $\range\Gamma$ (see \eqref{state_cov_mat}).
    By construction, we have
    \begin{equation*}
    	\trace\Sigma = \sum_{k} |c_k| \|G(e^{i\theta_k})\| = \tau.
    \end{equation*}
%    $\trace\Sigma = \sum_{k} |c_k| \|G(e^{i\theta_k})\| = \tau$. 
    Moreover, it is not difficult to verify the relation
	\begin{equation}\label{decomp_mat_primal_constraint}
	\bmat \tau&\xb^*\\ \xb&\Sigma \emat = \sum_{k} \frac{|c_k|}{\|G(e^{i\theta_k})\|} \bmat u_k^*\\ G(e^{i\theta_k}) \emat \bmat u_k & G^*(e^{i\theta_k}) \emat \geq 0.
	\end{equation}
	Therefore, $(\tau,\Sigma)$ is feasible for the optimization problem \eqref{AN_semidef_program}, and it must hold that $p \leq \frac{1}{2} (\tau + \trace\Sigma) = \sum_{k}|c_k| \|G(e^{i\theta_k})\|$ because $p$ is the minimum value. Since the inequality holds for any atomic decomposition of $\xb$, it must hold for the infimum, i.e., $p\leq\|\xb\|_\Acal$.

	Next we show the opposite inequality $\|\xb\|_\Acal\leq p$. Let a minimizer of \eqref{AN_semidef_program} be $(\hat{\tau}, \hat{\Sigma})$. Given the linear matrix inequality (LMI) constraint \eqref{LMI_constraint}, we have $\hat{\Sigma}\geq0$. Thus in view of Theorem~\ref{thm:C-F_type_decomp}, we can write down the unique C--F-type decomposition $\hat{\Sigma} = \sum_{k=1}^{\hat{r}} \hat{\rho}_k G(e^{i\hat{\theta}_k}) G^*(e^{i\hat{\theta}_k})$ with positive $\hat{\rho}_k$'s. According to the theory of the generalized Schur complement \citep[{Theorem 1.19 iv)}]{zhang2006schur}, we have $\xb\in\range \hat{\Sigma}$, which means that there exist complex numbers $\{\hat{c}_k\}$ such that
	\begin{equation}\label{atomic_decomp_xb}
	\xb = \sum_{k=1}^{\hat{r}} G(e^{i\hat{\theta}_k}) \hat{c}_k =: \Gb(\hat{\thetab})\hat{\cb}.
	\end{equation}
	The latter notation is just a matrix-vector product for the sum such that $\hat{\thetab} = (\hat\theta_1,\dots,\hat{\theta}_{\hat{r}})$ and $\Gb(\hat{\thetab})$ has columns $\{G(e^{i\hat{\theta}_k})\}$. 
	Appealing to the Schur complement again, it holds that
	\begin{equation}\label{inequal_b}
	\begin{split}
	\hat{\tau} & \geq \xb^* \hat{\Sigma}^\dagger \xb \\
	& = \hat{\cb}^* \Gb^*(\hat{\thetab}) \left[ \Gb(\hat{\thetab}) \hat{\Rb} \Gb^*(\hat{\thetab}) \right]^\dagger \Gb(\hat{\thetab})\hat{\cb} \\
	& = \hat{\cb}^* \Gb^*(\hat{\thetab}) [\Gb^*(\hat{\thetab})]^\dagger \hat{\Rb}^{-1} [\Gb(\hat{\thetab})]^\dagger \Gb(\hat{\thetab}) \hat{\cb} \\
	& = \hat{\cb}^* \hat{\Rb}^{-1} \hat{\cb} = \sum_{k=1}^{\hat{r}} \frac{|\hat{c}_k|^2}{\hat{\rho}_k}
	\end{split}
	\end{equation}
	where $^\dagger$ denotes the Moore-Penrose pseudoinverse, $\hat{\Rb}:=\diag\{\hat{\rho}_1,\dots,\hat{\rho}_{\hat{r}}\}$ is an invertible diagonal matrix, and we have used the fact that $\Gb(\hat{\thetab})$ has linearly independent columns. 
	Finally, we arrive at the chain of inequalities
	\begin{equation}\label{p_geq_x_A}
	\begin{split}
	p & =\frac{1}{2} (\hat{\tau} + \trace\hat{\Sigma}) \\
	& \geq \frac{1}{2} \sum_{k=1}^{\hat{r}} \left( \frac{|\hat{c}_k|^2}{\hat{\rho}_k} + \hat{\rho}_k \|G(e^{i\hat{\theta}_k})\|^2 \right) \\
	& \geq \sum_{k=1}^{\hat{r}} |\hat{c}_k| \|G(e^{i\hat{\theta}_k})\| \geq \|\xb\|_\Acal,
	\end{split}
	\end{equation}
	where we have used \eqref{inequal_b} and the definition of the atomic norm. \pfend
\end{pf}

\begin{remark}
	Due to the LMI constraint \eqref{LMI_constraint}, $\Sigma\in\range\Gamma$ can be interpreted as the signal covariance matrix $\E (\xb\xb^*)$ under suitable statistical assumptions on the amplitudes $\{a_k\}$ in \eqref{signal_model} which also appear in \eqref{filtered_amp}. In fact, subspace methods for frequency estimation are based on a similar idea, see \cite{stoica2005spectral,georgiou2000signal,amini2006tunable}. However, the ANM approach differs from subspace methods which \emph{estimate} the covariance matrix from the signal measurements, in that $\Sigma$ is constructed via the solution of the SDP \eqref{AN_semidef_program} which explicitly enforces low-rankness via the \emph{nuclear norm} ($\trace\Sigma$ in \eqref{obj_noiseless}) and the membership in the linear subspace $\range\Gamma$.
\end{remark}

Theorem~\ref{thm_AN_noiseless} suggests a way of doing frequency estimation (in the noiseless case) by first solving the optimization problem \eqref{AN_semidef_program}. Then given the optimal $\hat{\Sigma}$, the unknown frequencies $\{\hat\theta_k\}$ can be recovered via computing its C--F-type decomposition in the sense of Theorem~\ref{thm:C-F_type_decomp} which is illustrated in Example~\ref{ex_C-F-type_decomp}.

\begin{remark}
    Our Theorem~\ref{thm_AN_noiseless} does not follow from the general theory of the atomic norm in \citet{chandrasekaran2012convex} where the SDP representation is developed for atomic sets $\Acal$ that are \emph{real algebraic varieties}, i.e., real solutions to a system of polynomial equations. The reason is given as follows.
    It is only known that for our frequency estimation problem, the atomic set $\Acal$ in \eqref{dictionary} is a real algebraic variety \emph{in the special case of a delay filter bank} (Example~\ref{ex_delay_filt_bank}). In fact, the convex hull of such $\Acal$ is called a \emph{Carath\'{e}odory orbitope} which is studied in detail in \cite{sanyal2011orbitopes}. In the general case of a rational G-filter, however, it is not known whether $\Acal$ in \eqref{dictionary} is a real algebraic variety. Therefore, the results in \citet[Section~4]{chandrasekaran2012convex} does not seem directly applicable.
    In addition, the atomic decomposition \eqref{atomic_decomp_xb} can be effectively computed in a unique fashion all thanks to the C--F-type decomposition (Theorem~\ref{thm:C-F_type_decomp}). It must be noted that this latter feature is not common in the general theory of the atomic norm, that is, the general theory of semidefinite representation does not provide any means for the practical construction of an atomic decomposition on a specific atomic set, see the discussion in \citet[p.~838]{chandrasekaran2012convex}.
\end{remark}

\section{Dual problem}\label{sec:dual}

Every norm has an associated dual norm. The dual of the atomic norm \eqref{atomic_norm} plays a role in the dual optimization problem of the trivial primal
\begin{subequations}\label{AN_primal}
	\begin{align}
	& \underset{\zb\in\Cbb^n}{\text{minimize}}
	& & \|\zb\|_\Acal \label{primal_obj} \\
	& \text{subject to}
	& & \zb=\xb, \label{primal_constraint} %\\
	\end{align}
\end{subequations}
and the solution to the dual problem can also be exploited to extract the frequencies $\{\theta_k\}$ from the signal vector $\xb$. 

Take an arbitrary $\qb\in \Cbb^n$. According to e.g., \citet[Appendix A.1.6]{bv_cvxbook}, the dual norm of $\|\cdot\|_\Acal$ is defined as
\begin{equation}\label{atomic_dual_norm}
\begin{aligned}
\|\qb\|_\Acal^* & := \sup_{\|\zb\|_\Acal\leq 1} \innerprod{\qb}{\zb}_{\Rbb} \\
 & = \sup_{\|\zb\|_\Acal\leq 1} \sum_{k} \Re\{c_k^* G^*(e^{i\theta_k}) \qb\} \\
 & = \sup_{\|\zb\|_\Acal\leq 1} \sum_{k} |c_k| \|G(e^{i\theta_k})\| \, \Re\left\{e^{-i\phi_k}\frac{G^*(e^{i\theta_k})}{\|G(e^{i\theta_k})\|} \qb\right\} \\
 & = \max_{\phi,\, \theta\in \Ical} \Re\left\{e^{-i\phi}\frac{G^*(e^{i\theta})}{\|G(e^{i\theta})\|} \qb\right\} \\
 & =  \max_{\theta\in \Ical} \frac{|G^*(e^{i\theta}) \qb|}{\|G(e^{i\theta})\|},
\end{aligned}
\end{equation}
where $\phi_k\in\Ical$ is such that $c_k=|c_k|e^{i\phi_k}$,
and the fourth equality can be understood via an argument similar to that a solution of a linear program can only be obtained at vertices of the feasible polytope.
Define a rational function 
\begin{equation}\label{Q_def}
	Q(\theta):=G^*(e^{i\theta})\qb.
\end{equation} 
Then the above computation implies that the dual atomic norm $\|\qb\|_\Acal^*$ is equal to the maximum of the modulus of $Q(\theta)$ normalized by the ``weight'' of the atom $\|G(e^{i\theta})\|$.
The dual problem of \eqref{AN_primal} is
\begin{subequations}\label{AN_dual}
	\begin{align}
	& \underset{\qb\in\Cbb^n}{\text{maximize}}
	& & \innerprod{\qb}{\xb}_{\Rbb} \label{dual_obj} \\
	& \text{subject to}
	& & \|\qb\|_\Acal^* \leq 1, \label{dual_constraint} %\\
	\end{align}
\end{subequations}
see \cite{bv_cvxbook}, where the constraint can equivalently be written as $|Q(\theta)|\leq \|G(e^{i\theta})\|$ for all $\theta\in\Ical$ in view of \eqref{atomic_dual_norm}.

Let $\qb$ be dual feasible. The primal problem has only equality constraints, so Slater's condition holds trivially, which implies strong duality \citep{bv_cvxbook}. In fact, the primal problem has only the trivial solution $\zb=\xb$. While the weak duality leads to the relation
\begin{equation}
\innerprod{\qb}{\zb}_{\Rbb} = \innerprod{\qb}{\xb}_{\Rbb} \leq \|\xb\|_\Acal,
\end{equation}
a consequence of the strong duality is that the equality above holds if and only if $\qb$ is dual optimal.
More precisely, we have the next proposition in which the set $\Theta:=\{\theta_k\}_{k=1}^m$ contains the unknown true frequencies.

\begin{proposition}\label{prop_dual_certif}
	If there exists a rational function $Q(\theta)=G^*(e^{i\theta})\qb$ such that
\begin{subequations}\label{dual_certif}
	\begin{align}
	Q(\theta_k) = \frac{c_k}{|c_k|}\|G(e^{i\theta_k})\|,\ & \forall \theta_k\in\Theta, \label{cond_interpola} \\
	\text{and}\ |Q(\theta)| < \|G(e^{i\theta})\|, \ & \forall \theta\notin\Theta, \label{cond_inequal}
	\end{align}
\end{subequations}
then the true atomic decomposition $\xb=\sum_{k=1}^m G(e^{i\theta_k})c_k$ 
%in \eqref{state_atomic_decomp} 
is the unique one that achieves the atomic norm $\|\xb\|_\Acal$.
\end{proposition}

\begin{pf}
	We notice first that any $\qb$ satisfying \eqref{cond_interpola} and \eqref{cond_inequal} is feasible for \eqref{AN_dual}. By the definition of the dual norm, we automatically have a H\"{o}lder-type inequality $\innerprod{\qb}{\xb}_{\Rbb}\leq \|\qb\|_\Acal^* \|\xb\|_\Acal\leq \|\xb\|_\Acal$ in which the second inequality follows from the constraint \eqref{dual_constraint}. In addition, we can carry out the following computation:
	\begin{equation}\label{dual_primal_obj_inequal}
	\begin{aligned}
	\innerprod{\qb}{\xb}_{\Rbb} & = \sum_{k=1}^m \innerprod{\qb}{G(e^{i\theta_k})c_k}_{\Rbb} \\
	& = \sum_{k=1}^m \Re\{c_k^* Q(\theta_k)\} \\
	& = \sum_{k=1}^m |c_k| \|G(e^{i\theta_k})\| \geq \|\xb\|_\Acal
	\end{aligned}
	\end{equation} 
	where we have used the atomic decomposition for $\xb$, the condition \eqref{cond_interpola}, and the definition of the atomic norm \eqref{atomic_norm}. Therefore, we have $\innerprod{\qb}{\xb}_{\Rbb}= \|\xb\|_\Acal$ which in view of strong duality, implies that $\qb$ is dual optimal.
	
	Now we can prove the uniqueness of the atomic decomposition that achieves the atomic norm.
	Suppose that $\xb = \sum_{k} G(e^{i\hat{\theta}_k}) \hat{c}_k$ is another atomic decomposition such that $\|\xb\|_\Acal=\sum_{k} |\hat c_k| \|G(e^{i \hat\theta_k})\|$. We can make similar computation with the $\qb$ in the premise of the proposition:
	\begin{equation}\label{dual_primal_obj_inequal_contradict}
	\begin{aligned}
	\innerprod{\qb}{\xb}_{\Rbb} & = \sum_{k} \Re\{\hat c_k^* Q(\hat \theta_k)\} \\
	 & = \sum_{\hat{\theta}_k\in\Theta} \Re\{\hat c_k^* Q(\hat \theta_k)\} + \sum_{\hat{\theta}_k\notin\Theta} \Re\{\hat c_k^* Q(\hat \theta_k)\}\\
	 & < \sum_{\hat{\theta}_k\in\Theta} |\hat c_k| \|G(e^{i\hat\theta_k})\| + \sum_{\hat{\theta}_k\notin\Theta} |\hat c_k| \|G(e^{i\hat\theta_k})\| \\
	 & = \|\xb\|_\Acal,
	\end{aligned}
	\end{equation} 
	where the strict inequality comes from the simple fact $\Re(z)\leq |z|$ for a complex number $z$, and the condition \eqref{cond_inequal} for $\hat\theta_k\notin\Theta$. This is a contradiction to \eqref{dual_primal_obj_inequal}, so it must happen that all $\hat{\theta}_k\in\Theta$. By a linear independence argument similar to the one in the proof of Theorem~\ref{thm:C-F_type_decomp} (which works whenever $m\leq n$), we can conclude that the two atomic decompositions $\xb=\sum_{k=1}^m G(e^{i\theta_k})c_k=\sum_{k=1}^m G(e^{i\theta_k})\hat c_k$ must be the same. 
	\pfend
\end{pf}

\begin{figure}
	\centering
	\includegraphics[width=0.35\textwidth]{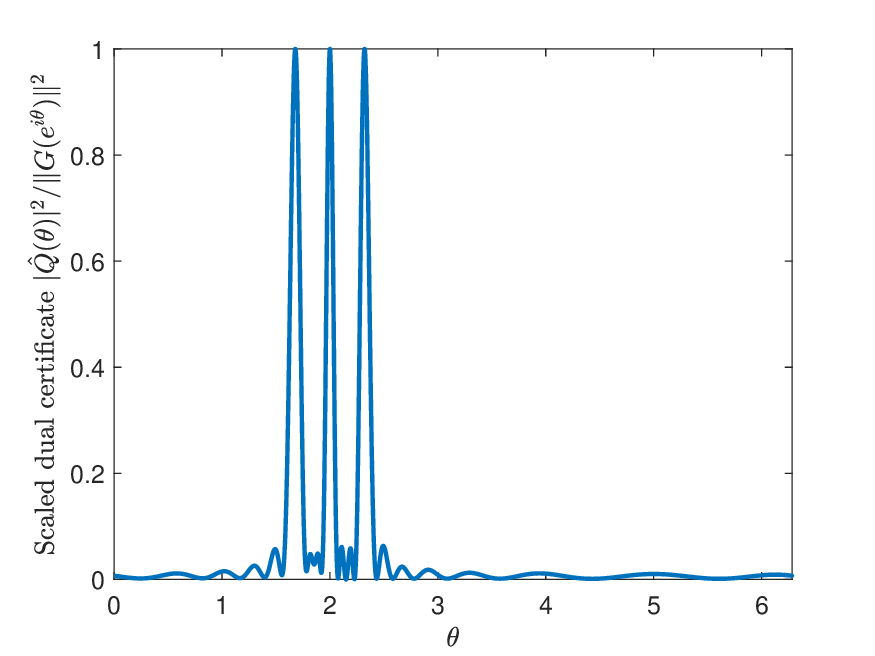}
	\caption{Function values of the scaled dual certificate $|\hat Q(\theta)|^2 / \|G(e^{i\theta})\|^2$ on an equidistant grid in $\Ical$ of size $10^4$. The noiseless signal $y(t)$ has length $L=98$ and consists of $m=3$ cisoids, and the true frequencies are $\theta_1=2-5(2\pi/L)$, $\theta_2=2$, and $\theta_3=2+5(2\pi/L)$ which correspond to the three maxima of the scaled dual certificate with a function value $1$.}\label{fig_scale_dual_certi_noiseless}
\end{figure}

The rational function $Q(\theta)$ satisfying the two properties \eqref{cond_interpola} and \eqref{cond_inequal} is called a \emph{dual certificate} as it certifies the optimality of the true atomic decomposition that constitutes the signal $\xb$. Moreover, it is possible to use the dual optimal solution $\hat{\qb}$ {and the corresponding rational function $\hat Q(\theta):=G^*(e^{i\theta})\hat\qb$} to locate the frequencies $\{\theta_k\}$. More precisely, one can evaluate the nonnegative rational function
\begin{equation}\label{ration_func_f}
\begin{aligned}
	\hat f(\theta) & := \|G(e^{i\theta})\|^2-|\hat Q(\theta)|^2 \\
	 & = G^*(e^{i\theta}) (I-\hat\qb\hat\qb^*) G(e^{i\theta})
\end{aligned}
\end{equation}
on the frequency interval $\Ical$, and identify the zeros, that is, frequencies such that $\hat f(\theta)=0$. Alternatively, we can scale the dual certificate as $|\hat Q(\theta)|^2 / \|G(e^{i\theta})\|^2$ so that it takes value in $[0, 1]$, and this is illustrated in Fig.~\ref{fig_scale_dual_certi_noiseless}
where the length of the noiseless signal $y(t)$ is $L=98$, the number of cisoids is $m=3$, the true frequencies are $\theta_1=2-5(2\pi/L)$, $\theta_2=2$, and $\theta_3=2+5(2\pi/L)$, and the function values are computed on an equidistant grid in $\Ical$ of size $10^4$. Clearly, the frequencies are found 
at the peak of the scaled dual certificate. A clustering idea similar to that in Example~\ref{ex_C-F-type_decomp} can be used to make the peak localization more stable.
 
We must point out that in general, the dual problem \eqref{AN_dual} may admit multiple solutions. However, it is not difficult to show that any dual optimal solution must be able to locate the frequencies in $\Theta$. To make this statement more precise, let us define $\hat{\Theta}:=\{\theta\in\Ical : \hat{f}(\theta) =0\}$ to be the set of frequencies recovered from $\hat{Q}$. Then we must have $\Theta\subset\hat{\Theta}$. To this end, suppose that the opposite is true, namely $\Theta\backslash\hat{\Theta}\neq\emptyset$. A consequence is
\begin{equation}
\begin{aligned}
\innerprod{\hat \qb}{\xb}_{\Rbb} & = \sum_{\theta_k\in\Theta\cap \hat{\Theta}} \Re\{c_k^* \hat Q(\theta_k)\} + \sum_{\theta_k\in \Theta\backslash\hat{\Theta}} \Re\{c_k^* \hat Q(\theta_k)\}\\
& < \sum_{\theta_k\in\Theta\cap \hat{\Theta}} |c_k| \|G(e^{i\theta_k})\| + \sum_{\theta_k\in \Theta\backslash\hat{\Theta}} |c_k| \|G(e^{i\theta_k})\| \\
& = \|\xb\|_\Acal
\end{aligned}
\end{equation} 
which is similar to \eqref{dual_primal_obj_inequal_contradict}. The strict inequality comes from the property $|\hat Q(\theta_k)| < \|G(e^{i\theta_k})\|$ for $\theta_k\in\Theta\backslash\hat{\Theta}$, and it is a contradiction to strong duality. Hence, the inclusion $\Theta\subset\hat{\Theta}$ holds.

It is still possible that $\hat{\Theta}$ contains spurious frequencies that are not present in the signal $\xb$. However, by exploiting the fact that the dual constraint \eqref{dual_constraint} can also be represented via a LMI, we can expect most SDP solvers to find good dual certificates. In order to see this, we first notice that the dual problem of \eqref{AN_semidef_program} is also a SDP:
\begin{subequations}\label{dual_AN_semidef_program}
	\begin{align}
	& \underset{\substack{\qb\in\Cbb^n,\\Y\in\ker\Gamma^*}}{\text{maximize}}
	& & \innerprod{\qb}{\xb}_{\Rbb} \\
	& \text{subject to}
	& & \bmat I+Y & \qb \\ \qb^* & 1 \emat \geq 0. \label{dual_LMI_constraint} %\\
	\end{align}
\end{subequations}

We remark that the LMI in \eqref{dual_LMI_constraint} together with the constraint $Y\in\ker\Gamma^*$ is equivalent to the dual norm constraint in \eqref{dual_constraint}, and this fact is related to the \emph{Gram matrix parametrization} for symmetric polynomials, see \cite{dumitrescu2017positive}. Practically, it is not necessary to solve the dual SDP explicitly because most solvers can return a dual optimal solution for free when solving the primal problem.
The next proposition gives conditions for the ideal case $\hat \Theta=\Theta$, and the proof needs an auxiliary lemma from the appendix.

\begin{proposition}[Exact recovery of the frequencies]\label{prop_dual_SDP}
	For the signal $\xb=\sum_{k=1}^m G(e^{i\theta_k})c_k$ with $m<n$,
	let $\Dcal:=\{(\hat \qb, \hat Y)\}$ be the set of optimal solutions to the dual SDP \eqref{dual_AN_semidef_program}. If there exists some $(\hat \qb, \hat Y)\in \Dcal$ such that the rational function $\hat Q(\theta)=G^*(e^{i\theta})\hat \qb$ satisfies the properties \eqref{cond_interpola} and \eqref{cond_inequal}, then the following statements hold.
	\begin{enumerate}
		\item Any $(\hat \qb, \hat Y)$ in $\relint \Dcal$ forms a strict complementary pair\footnote{The readers may refer to \cite{alizadeh1997complementarity} for strict complementarity in SDP.} with an optimal solution $(\hat \tau, \hat \Sigma)$ to the primal SDP \eqref{AN_semidef_program}. More precisely, we have
		\begin{equation}\label{rank_strict_complement}
		\rank \bmat \hat \tau&\xb^* \\ \xb&\hat \Sigma \emat=m,\ \rank \bmat I+\hat Y & \hat \qb \\ \hat \qb^* & 1 \emat =n+1-m,
		\end{equation} 
		where $\relint$ denotes the relative interior of a set.
		\item All $(\hat \qb, \hat Y)\in \relint\Dcal$ satisfies the conditions \eqref{cond_interpola} and \eqref{cond_inequal}.
		\item The dual central path converges to a point in $\relint\Dcal$.
	\end{enumerate}
\end{proposition}

\begin{pf}
	In order to prove Statements (1), let us consider first the specific $\hat \qb$ such that $\hat Q(\theta)=G^*(e^{i\theta})\hat \qb$ satisfies \eqref{cond_interpola} and \eqref{cond_inequal}.
	We notice that the rational function defined in \eqref{ration_func_f} is nonnegative and equal to zero at $\theta_k\in \Theta$. 
Take $R=I-\hat\qb\hat\qb^*$ in Lemma~\ref{lem_mat_repres_rat_func} in the appendix, and we know that there exist $\hat Y\in\ker\Gamma^*$ and a positive semidefinite matrix $\hat H$ of rank $n-m$ such that $I-\hat\qb\hat\qb^*+\hat Y=\hat H$.
This implies that the specific pair $(\hat{\qb}, \hat Y)$ is dual feasible and indeed dual optimal: the feasibility comes from the fact that the Schur complement of the matrix in \eqref{dual_LMI_constraint} with respect to the lower-right entry $1$ is just $I+\hat Y-\hat\qb\hat\qb^*\geq 0$, and the optimality comes from the first part of the proof of Proposition~\ref{prop_dual_certif}. Moreover, we have
\begin{equation}
\begin{aligned}
\rank \bmat I+\hat Y & \hat \qb \\ \hat \qb^* & 1 \emat & = \rank \bmat \hat H+\hat{\qb}\hat{\qb}^* & \hat \qb \\ \hat \qb^* & 1 \emat \\
 & = \rank \bmat \hat H & \zerob \\ \hat \qb^* & 1 \emat = n-m+1,
\end{aligned}
\end{equation}
where the second equality comes from block elimination and the last equality follows from the rank-nullity theorem.
According to Proposition~\ref{prop_dual_certif}, the atomic decomposition $\xb=\sum_{k=1}^m G(e^{i\theta_k})c_k$ is the unique one that achieves the atomic norm $\|\xb\|_\Acal$. Meanwhile, the optimal variables $(\hat{\tau},\hat{\Sigma})$ of the primal SDP \eqref{AN_semidef_program} can be constructed via the formulas \eqref{tau_def} and \eqref{Sigma_def}. In addition, the relation \eqref{decomp_mat_primal_constraint} holds, and it implies the first rank equality in \eqref{rank_strict_complement}. Therefore, the specific $(\hat{\qb}, \hat Y)$ and $(\hat{\tau},\hat{\Sigma})$ form a strict complementary pair.
The statement for any $(\tilde{\qb}, \tilde Y)$ in the relative interior of $\Dcal$ is a consequence of Lemma 3.1 in \cite{goldfarb1998interior} which claims that all matrices of the form $\bmat I+\tilde Y & \tilde \qb \\ \tilde \qb^* & 1 \emat$ such that $(\tilde{\qb}, \tilde Y)\in\relint\Dcal$ share the same column space, and thus have rank $n+1-m$. The latter means that such $(\tilde{\qb}, \tilde Y)$ also forms a strict complementary pair with $(\hat{\tau}, \hat{\Sigma})$.

Next we work on Statement (2). 
For each $k=1, \dots, m$, let us check the quadratic form
\begin{equation}
	\begin{aligned}
		 & \bmat G^*(e^{i\theta_k}) & \, -u_k\emat \bmat I+\hat Y & \hat \qb \\ \hat \qb^* & 1 \emat \bmat G(e^{i\theta_k}) \\ -u_k^*\emat \\
		= & \|G(e^{i\theta_k})\|^2 - 2\Re\left[ \hat Q(\theta_k) u_k^* \right] + |u_k|^2 = 0,
	\end{aligned}
\end{equation}
where, the quantity $u_k$ was defined in \eqref{u_k_def} which is also the right hand side of \eqref{cond_interpola}, and we have used the fact that $G^*(e^{i\theta}) \hat Y G(e^{i\theta})\equiv 0$, the definition of $Q$ in \eqref{Q_def}, and the condition \eqref{cond_interpola}. The above formula implies that
\begin{equation}
	\bmat G(e^{i\theta_k}) \\ -u_k^*\emat \in \ker \bmat I+\hat Y & \hat \qb \\ \hat \qb^* & 1 \emat,\ k=1, \dots, m.
\end{equation}
Since these $m$ vectors are linearly independent, it follows from the claim of Statement (1) that they form a basis of $\ker \bmat I+\hat Y & \hat \qb \\ \hat \qb^* & 1 \emat$. Referring again to \citet[Lemma~3.1]{goldfarb1998interior}, we know that the same $m$ vectors also form a basis of $\ker \bmat I+\tilde Y & \tilde \qb \\ \tilde \qb^* & 1 \emat$ for any $(\tilde{\qb}, \tilde{Y})\in\relint\Dcal$, that is,
\begin{equation}\label{quadrat_form_q_Y_tilde}
	\begin{aligned}
		& \bmat G^*(e^{i\theta_k}) & \, -u_k\emat 
		\bmat I+\tilde Y & \tilde \qb \\ \tilde \qb^* & 1 \emat 
		\bmat G(e^{i\theta_k}) \\ -u_k^*\emat \\
		= & 2\left( |u_k|^2 - \Re\left[ \tilde Q(\theta_k) u_k^* \right] \right) = 0.
	\end{aligned}
\end{equation}
The LMI constraint \eqref{dual_LMI_constraint} implies that $|\tilde Q(\theta)| \leq \|G(e^{i\theta})\|$, which leads to the inequalities
$$\Re\left[ \tilde Q(\theta_k) u_k^* \right] \leq |\tilde Q(\theta_k)| |u_k| \leq \|G(e^{i\theta_k})\| |u_k| = |u_k|^2.$$
In view of \eqref{quadrat_form_q_Y_tilde}, both inequalities above must hold with equalities. Consequently, it must happen that
%\begin{equation}
	$\tilde{Q}(\theta_k) = u_k$
%\end{equation}
which is precisely \eqref{cond_interpola}. Now suppose that there is some other $\tilde{\theta}\notin\Theta$ such that $|\tilde{Q}(\tilde{\theta})|=\|G(e^{i\tilde\theta})\|$. 
Then one can check through similar computations to \eqref{quadrat_form_q_Y_tilde} that
%\begin{equation}
%	\bmat G(e^{i\tilde\theta}) \\ -\tilde{Q}(\tilde{\theta})^* \emat \in \ker \bmat I+\tilde Y & \tilde \qb \\ \tilde \qb^* & 1 \emat \implies \dim \ker \bmat I+\tilde Y & \tilde \qb \\ \tilde \qb^* & 1 \emat \geq m+1,
%\end{equation}
\begin{equation}
\begin{aligned}
\bmat G(e^{i\tilde\theta}) \\ -\tilde{Q}(\tilde{\theta})^* \emat \in \ker \bmat I+\tilde Y & \tilde \qb \\ \tilde \qb^* & 1 \emat \implies \\
 \dim \ker \bmat I+\tilde Y & \tilde \qb \\ \tilde \qb^* & 1 \emat \geq m+1,
\end{aligned}
\end{equation}
where the condition $m<n$ plays a role in the linear independence. This is a contradiction with Statement (1). Therefore, we must have $|\tilde{Q}(\tilde{\theta})|<\|G(e^{i\tilde\theta})\|$ for all $\tilde{\theta}\notin\Theta$ which is \eqref{cond_inequal}.

Finally, given the fact that strictly complementary optimal solutions exist for the primal and dual SDPs, Statement (3) follows from standard results in SDP, see \citet[Lemma~3.4]{luo1998superlinear} 
and also \cite{de1997initialization,halicka2002convergence}.
	\pfend
\end{pf}

\begin{remark}\label{rem_exist_dual_certif}
    It remains a nontrivial issue to establish the existence of  a vector $\qb$ and the corresponding rational function $Q(\theta)$ which satisfies the conditions of interpolation \eqref{cond_interpola} and maxima in modulus \eqref{cond_inequal}.
    Previous works on ANM (e.g., \cite{tang2013compressed}) rely on mathematical techniques in \cite{candes2014towards} that center around the \emph{Fej\'er kernel} for the \emph{explicit construction} of a dual polynomial that meets the required  conditions. However, it seems difficult to extend their idea to our more general setup with a G-filter where instead of polynomials, we have rational functions.
\end{remark}

\section{Extension to the noisy case}\label{sec:noisy}

The general case of frequency estimation with noise can also be handled in a way similar to \eqref{AN_semidef_program}.
Referring to \eqref{state_atomic_decomp}, in the noisy case we have
\begin{equation}\label{state_plus_noise}
\begin{aligned}
\tilde\xb(t) & = G(z) [s(t)+w(t)] \\
& = \int_{\Ical} G(e^{i\theta}) e^{i\theta t} \left[ \d\hat{s}(\theta) + \d\hat{w}(\theta) \right] \\
& = \sum_{k=1}^m G(e^{i\theta_k}) c_k(t) + \tilde{\wb}(t)
\end{aligned}
\end{equation}
where, $c_k(t)$ is the filtered amplitude in \eqref{filtered_amp}, and the filtered noise vector $\tilde{\wb}(t)$ has a matricial power spectral density $\Phi_{\tilde{\wb}}(\theta)=G(e^{i\theta}) \Phi_w(\theta) G^*(e^{i\theta})$ in which $\Phi_w(\theta)$ is the spectral density of the input noise $w(t)$.

Again we assume that we have access to the output vector $\tilde\xb(t)$ at only one time instance $t$.
In this case, we drop the dependence on $t$ and write the measurement as $\tilde{\xb}\in\Cbb^n$.
Now let the signal part in $\tilde\xb$ be $\xb$.
The latter object corresponds to the cisoidal signal $s(t)=\sum_{k=1}^{m} a_k e^{i\theta_k t}$ in $y(t)$. Using $\|\xb\|_{\Acal}$ as a \emph{regularization} term, we set up the optimization problem
\begin{equation}\label{ANM_noisy}
	\underset{\xb \in \Cbb^n}{\text{minimize}} \quad
	 \frac{1}{2} \|\tilde\xb-\xb\|^2 + 
	\lambda \|\xb\|_{\Acal},
\end{equation} 
where $\lambda>0$ is a regularization parameter. In view of Theorem~\ref{thm_AN_noiseless}, such a problem admits the following SDP description
\begin{subequations}\label{noisy_SDP}
	\begin{align}
	& \underset{\substack{\tau\in\Rbb,\ \xb \in \Cbb^n\\ \Sigma\in\range\Gamma}}{\text{minimize}}
	& & \frac{1}{2} \|\tilde\xb-\xb\|^2 + \lambda (\tau + \trace\Sigma) \label{obj_noisy} \\
	& \text{subject to}
	& & \eqref{LMI_constraint}.
	\end{align}
\end{subequations}
The estimate of the frequencies is then computed from the C--F-type decomposition of the optimal $\hat\Sigma$.

A dual formulation is also available in this case. Indeed, {inspired by} Lemma 2 in \cite{bhaskar2013atomic}, the dual SDP is written as
\begin{subequations}\label{noisy_SDP_dual}
	\begin{align}
	& \underset{\substack{\qb \in \Cbb^n\\ Y\in\ker\Gamma^*}}{\text{maximize}}
	& & \innerprod{\qb}{\tilde{\xb}}_\Rbb - \frac{1}{2} \|\qb\|^2 \label{obj_noisy_dual} \\
	& \text{subject to}
	& & \bmat \lambda I+Y & \qb \\ \qb^* & \lambda \emat \geq 0,
	\end{align}
\end{subequations}
and the frequencies can alternatively be estimated by identifying the zeros of the rational function
\begin{equation}\label{ration_func_f_noisy}
\hat g(\theta) = G^*(e^{i\theta}) (\lambda I-\hat\qb\hat\qb^*/\lambda) G(e^{i\theta})
\end{equation}
in the frequency interval $\Ical$. {The ``noisy'' version of Proposition~\ref{prop_dual_certif} can be formulated in the style of \citet[Corollary~1, p.~5989]{bhaskar2013atomic}, but we choose to omit that for simplicity}.

{It remains to choose a suitable regularization parameter $\lambda$. According to \citet[Theorem~1]{bhaskar2013atomic}, one must choose $\lambda\geq \E\|\tilde\wb\|_{\Acal}^*$ in order to have a stable recovery of the signal component $\xb$. 
In the case of a delay filter bank, an estimate for $\E\|\tilde\wb\|_{\Acal}^*$ can be explicitly computed using \emph{Bernstein's inequality} for polynomials, see \citet[Subsec.~III-B]{bhaskar2013atomic}. However, as the atomic set changes to \eqref{dictionary} in the general case, it appears that new techniques are needed to estimate the expected dual atomic norm of $\tilde \wb$
and this is left for future work.
In our simulations to be presented in the next section, we take the \emph{heuristic} value
\begin{equation}\label{lambda_value_in_sims}
	\lambda =\frac{\sigma}{2} \sqrt{n\log n},
\end{equation}
which is the dominant term of the regularization parameter in the case of a delay filter bank of size $n$ under the additional assumption that the components of $\tilde\wb$ are i.i.d.~Gaussian with variance $\sigma^2$.

\section{Simulations}\label{sec:sims}

In this section, we perform numerical simulations for frequency estimation in the style of \citet[Sec.~V]{amini2006tunable}. More precisely, we redo Examples~2 and 3  in that paper with significantly shorter data sequences using the approach developed in the previous sections. 
{In addition, we make comparisons with alternative methods in the literature for frequency estimation.}
Some important details of implementation are described next.

%\noindent{\bf Construction of a G-filter.}
\subsection{Construction of a G-filter}\label{subsec:construct_G-filter}

Following \citet[Sec.~VII-F]{amini2006tunable}, we use G-filters with one repeated pole at $p=\rho e^{i\varphi}$ of multiplicity $n$. Such a filter should select a frequency band $[\theta_{\ell}, \theta_{u}]$ which represents our \emph{a priori} knowledge about the locations of the cisoids. By band selection, we mean that the filter gain $\|G(e^{i\theta})\|$ is relatively large inside the band. The parameters $(\rho, \varphi)$ are determined via a suboptimal procedure outlined in \citet[p.~2667]{amini2006tunable}\footnote{The optimal design of a G-filter with different poles can be a demanding task. However, this point should be worth investigating in a future work.}. Then the parameter pair $(A, \bb)$ of the G-filter can be constructed from the Jordan canonical form
\begin{equation}\label{filt_paras_Jordan}
	J =\bmat p & 1 & 0 & \cdots & 0 \\
	0 & p & 1 & \cdots & 0 \\
	\vdots & \vdots & \ddots & \ddots & \vdots \\
	0 & 0 & \cdots & p & 1 \\
	0 & 0 & \cdots & 0 & p\emat
	\quad \text{and}\quad
	\tilde\bb =\bmat 0 \\ 0 \\ \vdots \\ 0 \\ 1 \emat
\end{equation} 
via a suitable similarity transformation. Notice that $(J, \tilde{\bb})$ is reachable by construction. Apparently, the delay filter bank in Example~\ref{ex_delay_filt_bank} is a special case of $(J, \tilde{\bb})$ above with $p=0$. The need for a similarity transform comes from the normalization condition
\begin{equation}\label{filter_param_normaliz_cond}
	A A^* + \bb \bb^* = I,
\end{equation}
which amounts to saying that the discrete-time Lyapunov equation (DLE)
\begin{equation}\label{DLE}
	X - A X A^* = \bb \bb^*
\end{equation}
has the unique solution $X=I$. To this end, let $E$ be the solution to the DLE $E - J E J^* = \tilde\bb \tilde\bb^*$. Then we simply set $(A, \bb) = (E^{-1/2} J E^{1/2}, E^{-1/2}\tilde\bb)$ where $E^{1/2}$ can be any square root of $E$, e.g., the Cholesky factor or the Hermitian square root.

\begin{remark}
    The prior knowledge of the frequency band $[\theta_{\ell}, \theta_{u}]$ can come from various sources. For example, it can be deduced from a fast but not so accurate frequency estimator such as those based on the FFT \citep[p.~2666]{amini2006tunable}. Other cases include oversampling of the signal, or additional domain knowledge in specific applications such as radars and underwater acoustic communication, see e.g., \citet[p.~158]{yang2018frequency}. 
\end{remark}

%\noindent{\bf Filtering a finite-length signal $y(t)$.}
\subsection{Filtering a finite-length signal $y(t)$}\label{subsec:filter_finite_y}

We filter the cisoidal signal $y(t)$ with $t=0, 1, \dots, L-1$ to obtain the vectorial measurement $\xb(t)$ with an initial condition $\xb(-1)=\zerob$. In order to compensate the transient effect of the initial condition, we discard the first $L_{\srm}$ filtered samples, where $L_{\srm}$ is chosen such that $\|A^{L_{\srm}}\|<\varepsilon$ with a threshold $\varepsilon=10^{-3}$. As discussed in Remark~\ref{rem_input_filtering}, we shall emphasize the small-sample performance of our approach, and specifically, we assume that the number of measurements $L$ is so small that we can only obtain one single filtered output after the truncation, i.e., the quantity $L_{\xb}=L-L_{\srm}=1$ in \eqref{Sigma_sample_cov}.

%\noindent{\bf Choice of the regularization parameter $\lambda$ in the noisy case.} 
\subsection{Choice of the regularization parameter $\lambda$ in the noisy case}

{As discussed at the end of the previous section, we take the regularization parameter according to \eqref{lambda_value_in_sims} whose validity relies on the assumption that the components of the filtered noise $\tilde\wb$ is i.i.d.~Gaussian, namely $\tilde{\wb}\sim \mathcal{N}(\zerob, \sigma^2 I_n)$. We first explain how this Gaussianity assumption is satisfied in our setup. }
Assume that the noise $w(t)$ before filtering is i.i.d.~Gaussian with variance $\sigma^2$. Then obviously the filtered noise $\tilde\wb$ is a Gaussian random vector. The variance matrix of $\tilde{\wb}$ is just
\begin{equation}
	\E\tilde{\wb}\tilde{\wb}^* = \frac{\sigma^2}{2\pi} \int_{\Ical} G(e^{i\theta})G^*(e^{i\theta}) \d\theta = \sigma^2 \sum_{k=0}^{\infty} A^k \bb \bb^* (A^*)^k,
\end{equation}
and the latter series (without the multiplicative constant $\sigma^2$) is the unique solution to the DLE \eqref{DLE}
due to the fact that $\rho(A)<1$. 
Once the pair $(A, \bb)$ is suitably scaled as described in Subsection~\ref{subsec:construct_G-filter} 
%{\bf Construction of a G-filter} 
so that \eqref{filter_param_normaliz_cond} holds,
the components of $\tilde{\wb}$ are also i.i.d. Gaussian.

In practice, one still has to estimate the noise variance $\sigma^2$, and we follow the procedure in \cite{bhaskar2013atomic,Zhu-M2-LineSpec} which is briefly recalled as follows. Using the noisy measurements $y(t)$, we compute the \emph{standard biased covariance estimates} \citep{stoica2005spectral}
\begin{equation}
	\hat\sigma_y(k) = \frac{1}{L} \sum_{t=0}^{L-1-k} y(t+k)y(t)^*,\ k=0,1,\dots,\floor{L/3},
\end{equation}
where $\floor{x}$ represents the ``floor function'', i.e., the largest integer that is less than $x$. Then we form the Hermitian Toeplitz matrix $\hat{T}$ with $\bmat\hat\sigma_y(0)&\cdots&\hat\sigma_y(\floor{L/3})\emat^\top$ as the first column. The smallest $25\%$ of the eigenvalues of $\hat T$ are averaged, and the value is taken to be an estimate of the noise variance $\hat\sigma^2$. Such an estimate is statistically consistent due to the consistency of each $\hat\sigma_y(k)$ and the structure of the true $T$, see \citet[Subsec.~4.2.3, p.~150]{stoica2005spectral}.

%\noindent{\bf Characterization of the constraint $\Sigma\in\range\Gamma$.}
\subsection{Equality characterization of the constraint $\Sigma\in\range\Gamma$}
In principle, it is possible to explicitly compute a basis of $\range\Gamma$ by exploiting the characterization \eqref{charat_range_Gamma}.
However, a more computationally appealing way for the implementation can be found in \citet[Prop.~3.2]{ferrante2012maximum} in which it is shown that $\Sigma\in\range\Gamma$ is equivalent to the equality
\begin{equation}\label{equal_constraint}
	(I-\Pi_{\bb})(\Sigma-A\Sigma A^*)(I-\Pi_{\bb}) = O,
\end{equation}
where $\Pi_{\bb}:=\bb\bb^*/(\bb^*\bb)$ is the matrix of projection onto the line in the direction of $\bb$.

%\noindent{\bf Solving the SDPs.}
\subsection{Solving the SDPs}
The primal SDP \eqref{AN_semidef_program} and its noisy version \eqref{noisy_SDP} are solved using CVX, a package for specifying and solving convex programs \citep{cvx,gb08} which in turn calls SDPT3 \citep{toh1999sdpt3}.
The set membership constraint $\Sigma\in\range\Gamma$ is replaced by the equality constraint \eqref{equal_constraint}.
Once the optimal $\hat{\Sigma}$ is obtained,
the estimate of the frequencies $\hat{\thetab}=(\hat\theta_1,\dots,\hat\theta_{\hat{r}})$ is computed via the C--F-type decomposition as illustrated in Example~\ref{ex_C-F-type_decomp} where $\hat r$ is the \emph{numerical rank} of $\hat{\Sigma}$.
Here we emphasize that in this way, the ANM approach can \emph{automatically} detect the number $m$ of cisoids in the signal $y(t)$ by computing this numerical rank $\hat r$.
Such a feature is decisively different from subspace methods in which the number $m$ must be provided by other methods such as AIC or BIC \citep{stoica2005spectral}.
An alternative to frequency extraction is to identify the zeros of the rational function $\hat{f}(\theta)$ in \eqref{ration_func_f} (or  \eqref{ration_func_f_noisy} in the noisy case) with the optimal dual variable $\hat{\qb}$.

%\noindent{\bf Simulation results.} 
\subsection{Simulation results}

We present the simulation results in the form of four examples, 
and they are briefly described as follows.
\begin{enumerate}
    \item In Examples~\ref{ex_separa_freqs} and \ref{ex_high_resol}, we redo the simulations of Examples~2 and 3 in \cite{amini2006tunable} under nearly the same setup using our ANM approach. The differences include the signal length $L$ of $y(t)$, and the noise variance $\sigma^2$ because we do experiments with different signal-to-noise ratios (SNRs).

    \item In Example~\ref{ex_compare_FS}, we make comparisons between our approach, the standard ANM, and the frequency-selective  ANM (abbreviated as F-S ANM).

    \item In Example~\ref{ex_compare_subspace}, we compare the performances of our approach and two standard subspace methods which include MUSIC and ESPRIT.
\end{enumerate}
For each parameter configuration, we run a Monte Carlo simulation of $50$ repeated trials in order to evaluate the performance of an approach.

\begin{example}\label{ex_separa_freqs}
	There are three cisoids ($m=3$) in the signal $y$ of length $L=98$ such that the true frequencies are $\theta_1=\theta_0-5(2\pi/L)$, $\theta_2=\theta_0$, and $\theta_3=\theta_0+5(2\pi/L)$ where $\theta_0$ takes value from a finite candidate set $\{1.5, 1.7, 1.9, 2.1, 2.3, 2.5\}$. The amplitudes are $a_1=8e^{i\varphi_1}$, $a_2=4e^{i\varphi_2}$, and $a_3=2e^{i\varphi_3}$ where $\varphi_k, k=1,2,3$ are uniform random variables\footnote{The random phases is just a technical condition commonly assumed in the frequency estimation problem, see e.g., \cite{stoica2005spectral}.} in $[0, 2\pi]$. The frequencies are well separated in comparison with the resolution limit (i.e., the distance between two nearest frequencies {that can be distinguished) $\Delta_\text{FFT}:=2\pi/L$} of the FFT method.
	The SNR here is defined as $10\log_{10}(2^2/\sigma^2)$ dB using the smallest amplitude $|a_3|=2$. The candidate values for the SNR are $0, 3, 6, 9$ dB. The noise variance $\sigma^2$ is determined once the SNR is fixed, and is then used to generate complex Gaussian white noise $w(t)$ with $t=0, \dots, L-1$.
	
	We follow Subsection~\ref{subsec:construct_G-filter}
	%	{\bf Construction of a G-filter} 
	to design a G-filter of size $n=20$ which selects the frequency band $\Ical_1=[1.75, 2.25]$. According to \citet[Example~2]{amini2006tunable}, one should take the repeated pole $p=0.58e^{i2}$. 
	Notice that the candidate values $1.5$ and $2.5$ for $\theta_0$ fall outside the desired frequency band.
	The graph of the squared norm $\|G(e^{i\theta})\|^2$, which can be interpreted as the ``gain'' of the filter bank, versus the frequency $\theta$ is shown in Fig.~\ref{fig_Gfilter_gain}. Clearly, the curve is unimodal in $\Ical$: it has a peak at $\theta=2$ and decays on both sides of the peak. The mode of decay is almost symmetric with respect to $\theta=2$.
	In order to meet the tolerance level $\varepsilon=10^{-3}$ in Subsection~\ref{subsec:filter_finite_y},
	%	{\bf Filtering a finite-length signal $y(t)$}, 
	the number of filtered samples to be truncated is $L_{\srm}=97$ which implies that only \emph{one} sample is left for our ANM approach.

	For each candidate value of $\theta_0$, we run a Monte Carlo simulation to evaluate the performance of our approach. In Fig.~\ref{fig_probab_recov_three_freqs} we show the ``probability'' of successfully recovering the number of cisoids which is defined as $\#\{\text{trials} : \hat{r}=m=3\}/50$ where $\#\{\cdot\}$ denotes the cardinality of a set, and $\hat{r}$ is the numerical rank of the optimal $\hat{\Sigma}$ which is computed as follows. Let $\hat\lambda_k, k=1, \dots, n$ be the eigenvalues of $\hat{\Sigma}$ in nonincreasing order. Then $\hat r$ is equal to the first positive integer $k$ such that $\hat\lambda_{k+1}<10^{-3}$ or $\hat\lambda_k/\hat\lambda_{k+1}>10^3$. We observe that the probability is reasonably good (larger than $0.8$) when $1.5\leq \theta_0\leq 2.3$ and $\SNR\geq 3$ dB or when $1.5\leq \theta_0\leq 1.9$ and $\SNR=0$ dB. 
    {We notice that Fig.~\ref{fig_probab_recov_three_freqs} is not symmetric with respect to $\theta_0=2$. This behavior can be explained by the fact that the frequency component $\theta_3$ corresponds to the smallest amplitude $a_3$ in modulus, which means that the cisoid with $\theta_3$ becomes more and more difficult to detect when it moves away from the peak location of the gain of the G-filter (Fig.~\ref{fig_Gfilter_gain}) and thus gets less and less amplified.}
    Furthermore, for all the successful trials we compute the absolute error $\|\hat{\thetab}-\thetab\|$ where $\thetab=(\theta_1, \theta_2, \theta_3)$ is the true frequency vector, and these errors are depicted in Fig.~\ref{fig:sim_results_three_freqs} using the $\mathtt{boxplot}$. It is evident that the errors decrease as the SNR increases and the smallest errors happen at $\theta_0=1.7$ and $1.9$ which are close to the selected frequency band $\Ical_1$.

	\begin{figure}
		\centering
		\includegraphics[width=0.35\textwidth]{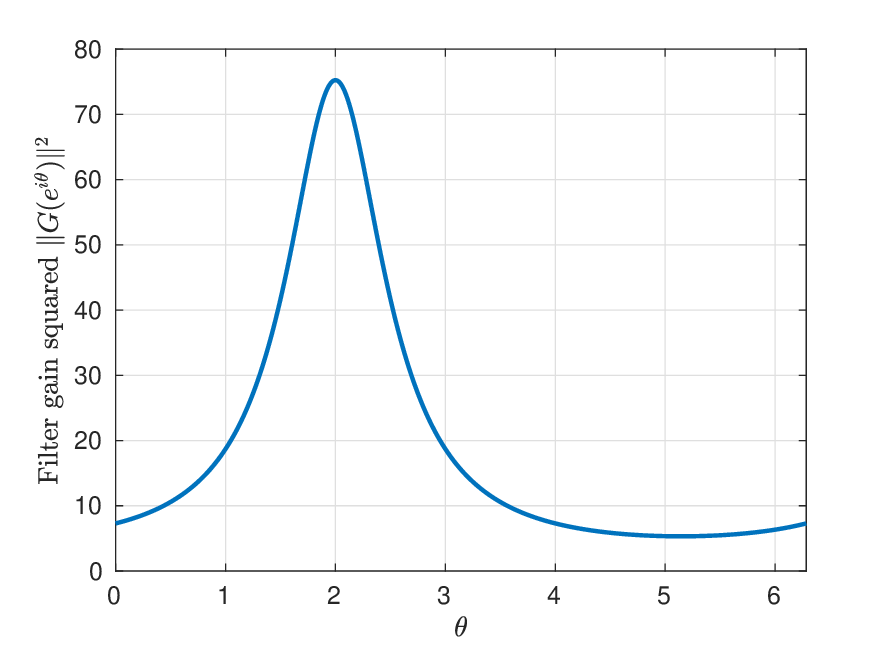}
		\caption{The squared ``gain'' $\|G(e^{i\theta})\|^2$ of a G-filter of size $n=20$ versus the frequency $\theta\in\Ical$. The filter parameters $(A, \bb)$ are constructed as per Subsec.~\ref{subsec:construct_G-filter} with a repeated pole $p=0.58e^{i2}$.}\label{fig_Gfilter_gain}
	\end{figure}

	\begin{figure}
		\centering
		\includegraphics[width=0.35\textwidth]{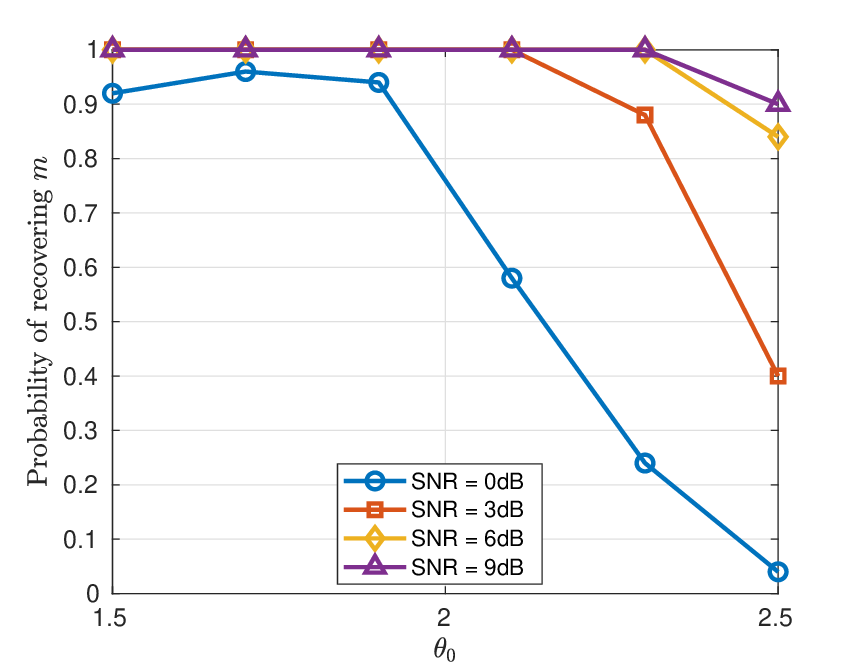}
		\caption{The ``probability'' of successfully recovering the number $m=3$ of cisoids versus $\theta_0$ under the SNRs $0, 3, 6, 9$ dB in Example~\ref{ex_separa_freqs}. Such a probability is defined by the number of successful instances in a Monte Carlo simulation divided by $50$. Notice that the markers represent computed values and lines have no meaning.}\label{fig_probab_recov_three_freqs}
	\end{figure}
	
	\begin{figure}
		%	\centering
		\begin{subfigure}[b]{.48\columnwidth}
			\includegraphics[width=\linewidth]{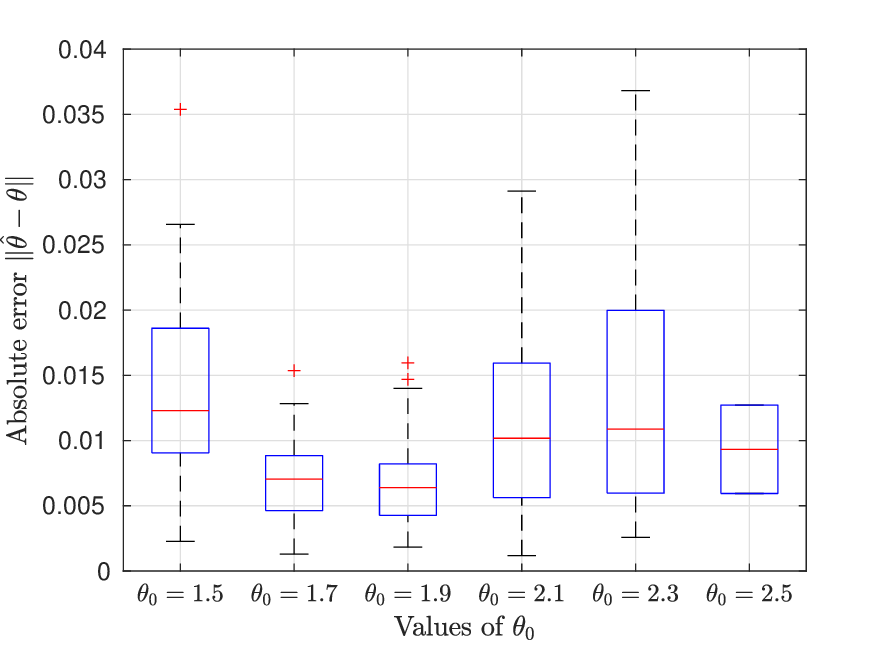}
			\caption{$\SNR=0$ dB.}
			\label{subfig:sims_separa_SNR1}
		\end{subfigure}
		\hfill
		\begin{subfigure}[b]{.48\columnwidth}
			\includegraphics[width=\linewidth]{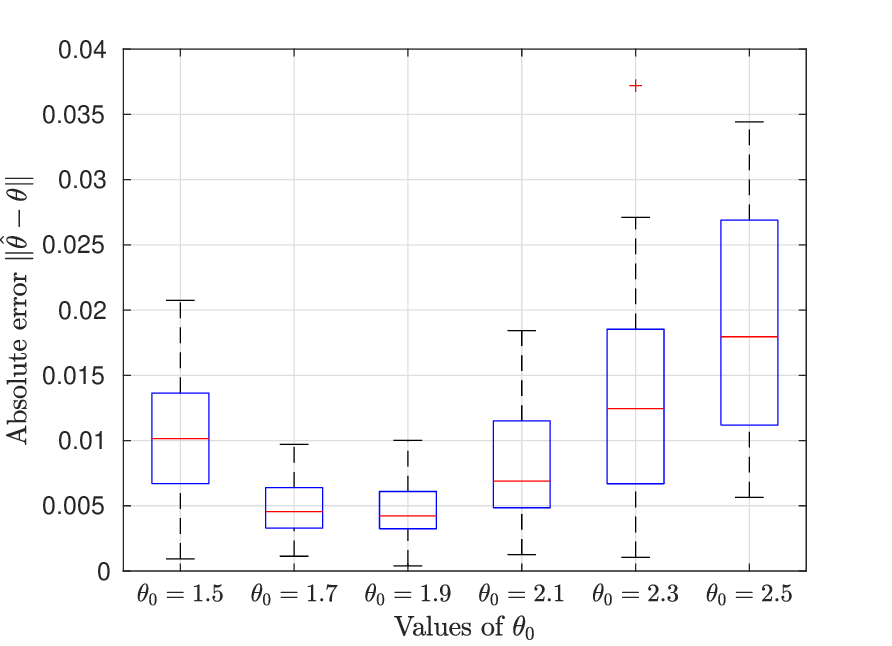}
			\caption{$\SNR=3$ dB.}
			\label{subfig:sims_separa_SNR2}
		\end{subfigure}
		
		\begin{subfigure}[b]{.48\columnwidth}
			\includegraphics[width=\linewidth]{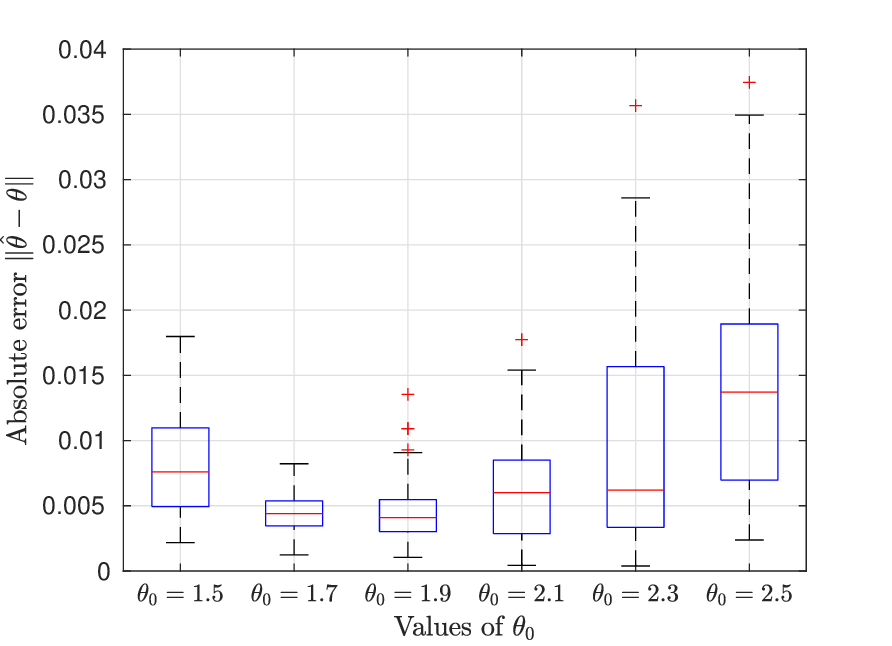}
			\caption{$\SNR=6$ dB.}
			\label{subfig:sims_separa_SNR3}
		\end{subfigure}
		\hfill
		\begin{subfigure}[b]{.48\columnwidth}
			\includegraphics[width=\linewidth]{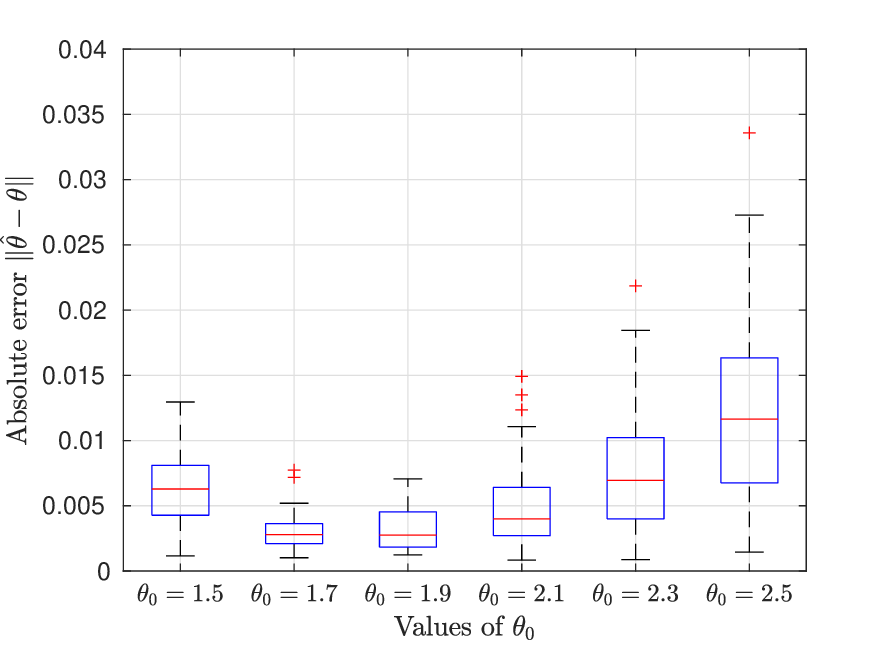}
			\caption{$\SNR=9$ dB.}
			\label{subfig:sims_separa_SNR4}
		\end{subfigure}
		
		\caption{Absolute errors $\|\hat{\thetab}-\thetab\|$ of frequency estimation in each Monte Carlo simulation under different SNRs as indicated by captions of the subfigures. Notice that the right-most box in Subfig.~(a) is not informative because it contains only two values.}
		\label{fig:sim_results_three_freqs}
	\end{figure}

\end{example}

\begin{example}\label{ex_high_resol}
	Here we study the \emph{high-resolution} property of our frequency estimator. More precisely, we consider the case with two cisoids ($m=2$)
	such that the true frequencies are $\theta_1=\theta_0-{\Delta_\text{FFT}/2}$, $\theta_2=\theta_0+{\Delta_\text{FFT}/2}$ and the amplitudes are $a_k=5e^{i\varphi_k}, k=1, 2$ with uniform random phases $\varphi_k$.
	The SNR now is defined as $10\log_{10}(5^2/\sigma^2)$ dB.
	The two frequencies are separated at a distance 
    equal to the resolution limit of the FFT method ({$\Delta_\text{FFT}$}).
	The rest parameters, including the signal length $L$, the candidate sets for $\theta_0$ and the SNR, and the size $n$ and the parameters $(A, \bb)$ of the G-filter, 
	are identical to those in Example~\ref{ex_separa_freqs}.

	Again we run a Monte Carlo simulation
    for each choice of $\theta_0$ and $\SNR$. The ``probabilities'' of successfully recovering the number $m=2$ are shown in Fig.~\ref{fig_probab_recov_two_freqs}.
	Then the absolute errors of frequency estimation in all the successful trials are computed and depicted in Fig.~\ref{fig:sim_results_two_freqs}. 
	We observe that the best performance occurs at $\theta_0=1.9$ or $2.1$ which falls in the selected band $\Ical_1$.
	Still the performance increases with the SNR. In general, resolving two closely located frequencies is a difficult task as we can see that the errors in Subfig.~\ref{fig:sim_results_two_freqs}(d) {are} larger than those in Subfig.~\ref{fig:sim_results_three_freqs}(d) with a factor of around $5$. 
	In addition, the design of the G-filter is also important because the performance of the ANM approach depends heavily on the point whether the frequency band of interest is properly selected. When the true frequencies fall outside the selected band, it is rather possible that the ANM approach returns bad estimates.

	\begin{figure}
		\centering
		\includegraphics[width=0.35\textwidth]{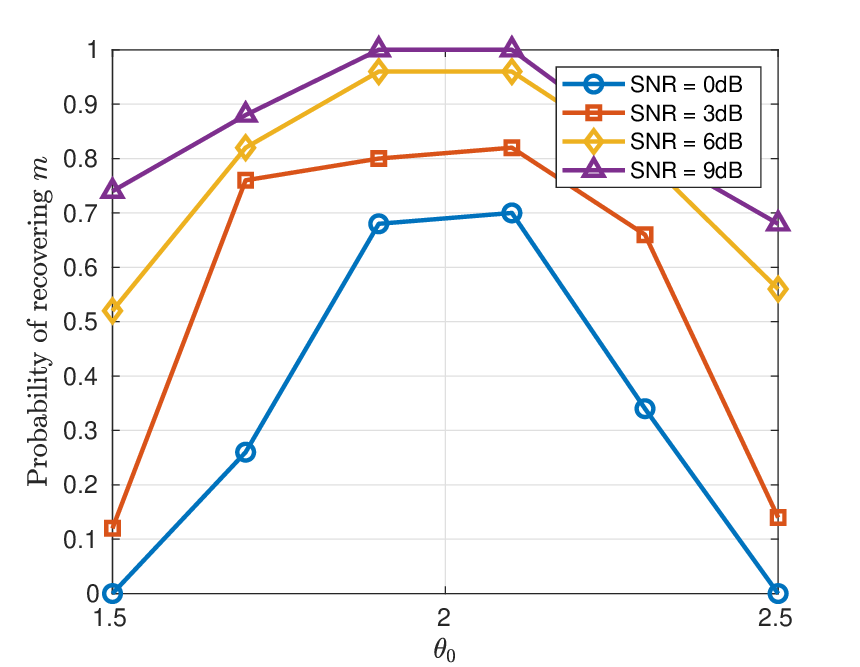}
		\caption{The ``probability'' of successfully recovering the number $m=2$ of cisoids versus $\theta_0$ under the SNRs $0, 3, 6, 9$ dB in Example~\ref{ex_high_resol}. Such a probability is defined by the number of successful instances in a Monte Carlo simulation divided by $50$. Notice that the markers represent computed values and lines have no meaning.}\label{fig_probab_recov_two_freqs}
	\end{figure}
	
	\begin{figure}
		%	\centering
		\begin{subfigure}[b]{.48\columnwidth}
			\includegraphics[width=\linewidth]{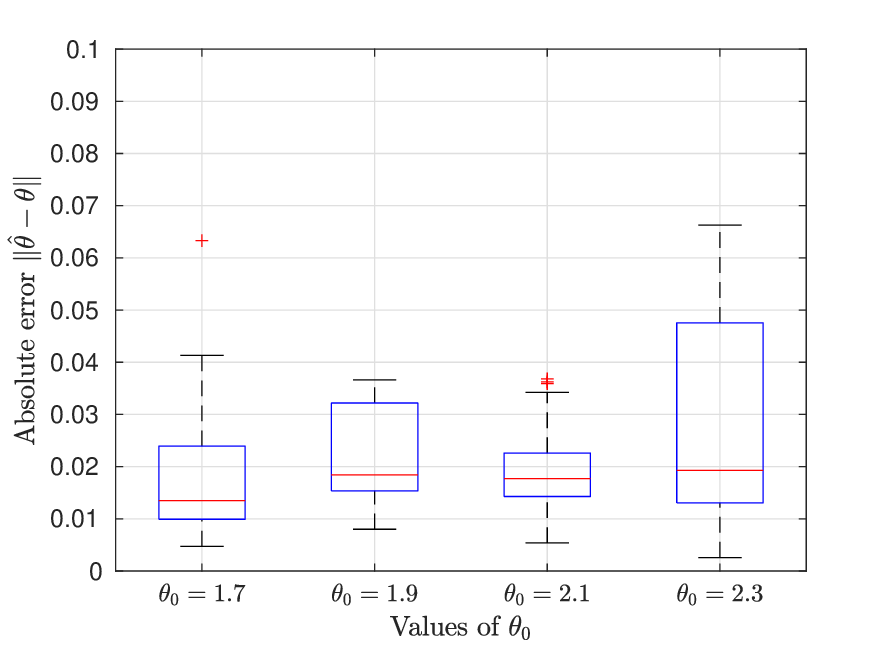}
			\caption{$\SNR=0$ dB.}
			\label{subfig:sims_high_res_SNR1}
		\end{subfigure}
		\hfill
		\begin{subfigure}[b]{.48\columnwidth}
			\includegraphics[width=\linewidth]{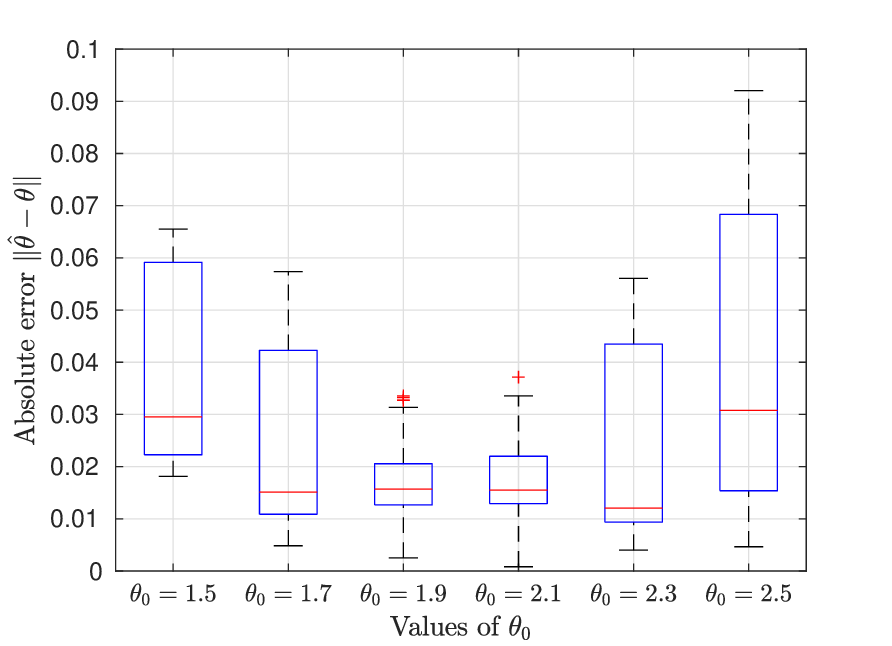}
			\caption{$\SNR=3$ dB.}
			\label{subfig:sims_high_res_SNR2}
		\end{subfigure}
		
		\begin{subfigure}[b]{.48\columnwidth}
			\includegraphics[width=\linewidth]{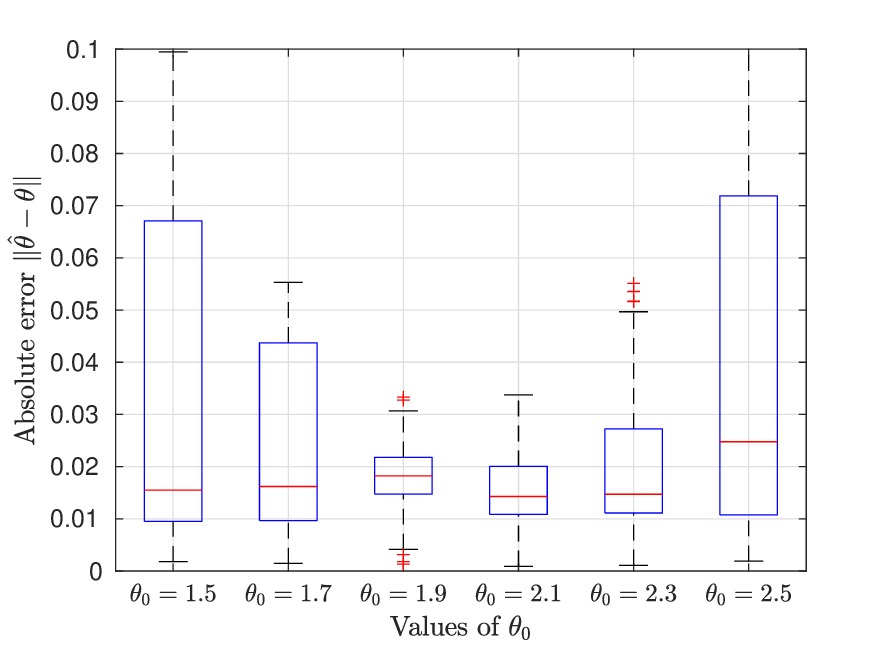}
			\caption{$\SNR=6$ dB.}
			\label{subfig:sims_high_res_SNR3}
		\end{subfigure}
		\hfill
		\begin{subfigure}[b]{.48\columnwidth}
			\includegraphics[width=\linewidth]{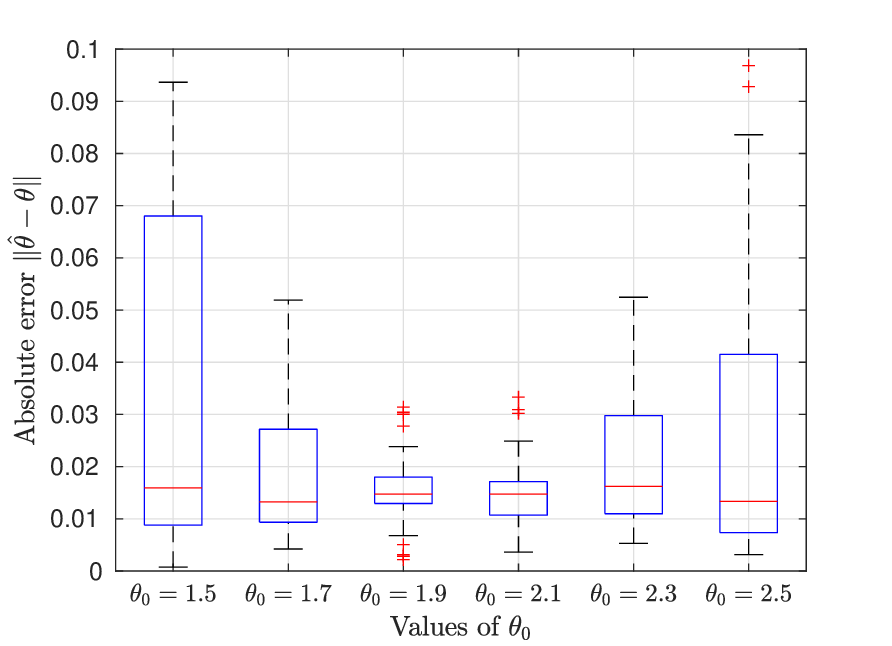}
			\caption{$\SNR=9$ dB.}
			\label{subfig:sims_high_res_SNR4}
		\end{subfigure}
		
		\caption{Absolute errors $\|\hat{\thetab}-\thetab\|$ of frequency estimation in each Monte Carlo simulation under different SNRs as indicated by captions of the subfigures. Notice that Subfig.~(a) contains only four boxes because the ANM approach cannot estimate the number $m=2$ of cisoids correctly in the cases of $\theta_0=1.5$ or $2.5$ as shown in the blue line of Fig.~\ref{fig_probab_recov_two_freqs}.}
		\label{fig:sim_results_two_freqs}
	\end{figure}

\end{example}

\begin{example}\label{ex_compare_FS}
	In this example, we compare our approach with  the standard ANM in \cite{bhaskar2013atomic} and the F-S ANM in \citet{yang2018frequency}. Here F-S stands for ``frequency-selective'' which means that prior knowledge is available saying that the frequencies are located in a specific band $\Ical_1$, much similar to our setup. The difference lies in the manner in which such prior knowledge is exploited: \cite{yang2018frequency} adapt the classic C--F decomposition and the standard ANM such that the frequency band is restricted to $\Ical_1$, while we use the G-filter to capture the band information.

    The parameters for simulations are given next.
		The number of cisoids  is $m=3$,
        and the length of  the signal $y$ is $L=98$. 
        The true frequencies are $\theta_1=\theta_0 - 2 \Delta_\text{FFT}$, $\theta_2=\theta_0$, and $\theta_3 = \theta_0 + 2\Delta_\text{FFT}$, where $\theta_0$ is selected from the candidate set $\{1.5, 1.6, \dots, 2.5\}$.
        That is, the frequency components are now spaced with a distance equal to twice of the resolution limit of the FFT method ($2\Delta_\text{FFT}$).
		The complex amplitudes are specified as $a_k = e^{i\varphi_k}$ with $\varphi_k \sim U[0, 2\pi]$ for $k=1, 2, 3$,
        and the $\SNR$ is defined as $10\log_{10}(1/\sigma^2)$ dB. The candidate values for the SNR are $-3, 0, 3, 6, 9$ dB. 
        The same G-filter used in the previous two examples is reused here which selects the band $\Ical_1=[1.75, 2.25]$.

		Again we run a Monte Carlo simulation for each choice of $\theta_0$ and $\SNR$.
        The  ``probabilities'' of successfully recovering the number $m=3$ are shown in Fig.~\ref{fig_ANMs_recov_compare} for our approach, the standard ANM, and the F-S ANM.
		We observe that the recovery probability of our approach (panel (a)) is close to $1$ when $1.8\leq \theta_0\leq 2.2$ and $\SNR\geq 3$ dB. In contrast, the recovery probability of the standard ANM (panel (b)) is below $0.7$ and the recovery probability of the F-S ANM (panel (c)) is only about $0.8$ even when $\SNR = 9$ dB.

	\begin{figure}
		\centering
		\begin{subfigure}[b]{.78\columnwidth}
			\includegraphics[width=\textwidth]{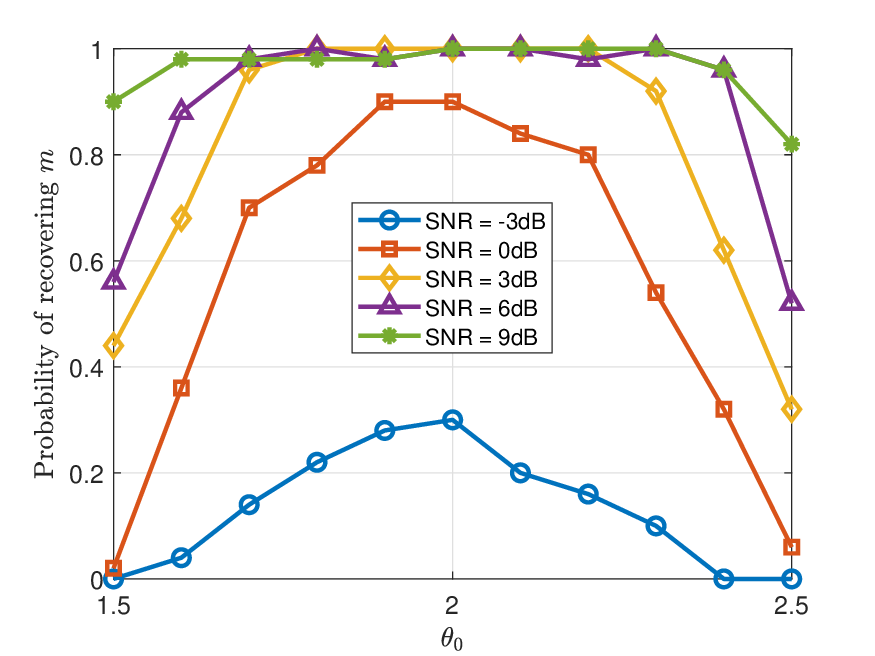}
			\caption{Our approach}
			\label{subfig:sims_Gfilt_SNR2}
		\end{subfigure}
		%	\hfill
		\begin{subfigure}[b]{.78\columnwidth}
			\includegraphics[width=\textwidth]{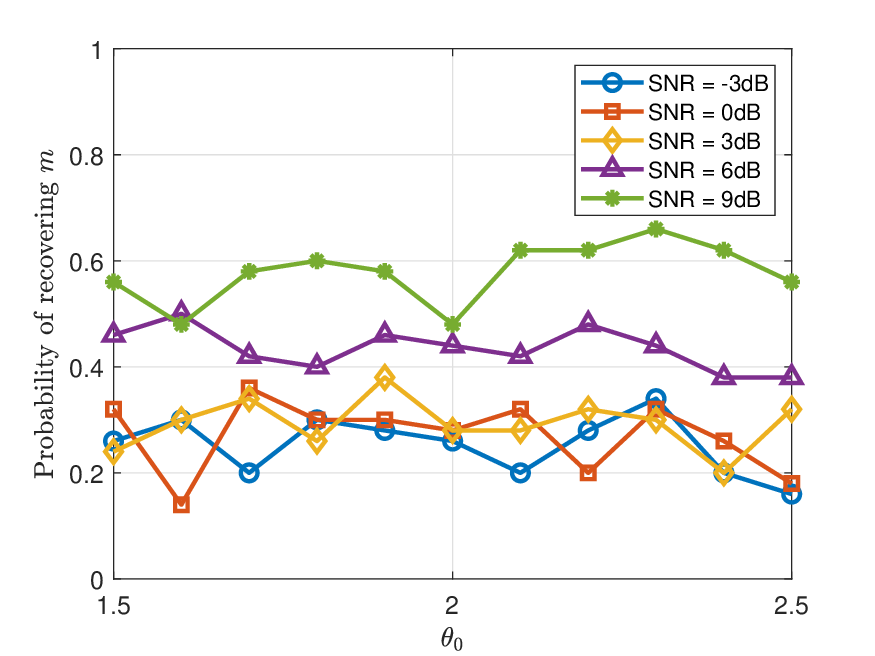}
			\caption{Standard ANM}
			\label{subfig:sims_Gfilt_SNR3}
		\end{subfigure}
		
		\begin{subfigure}[b]{.78\columnwidth}
			\includegraphics[width=\textwidth]{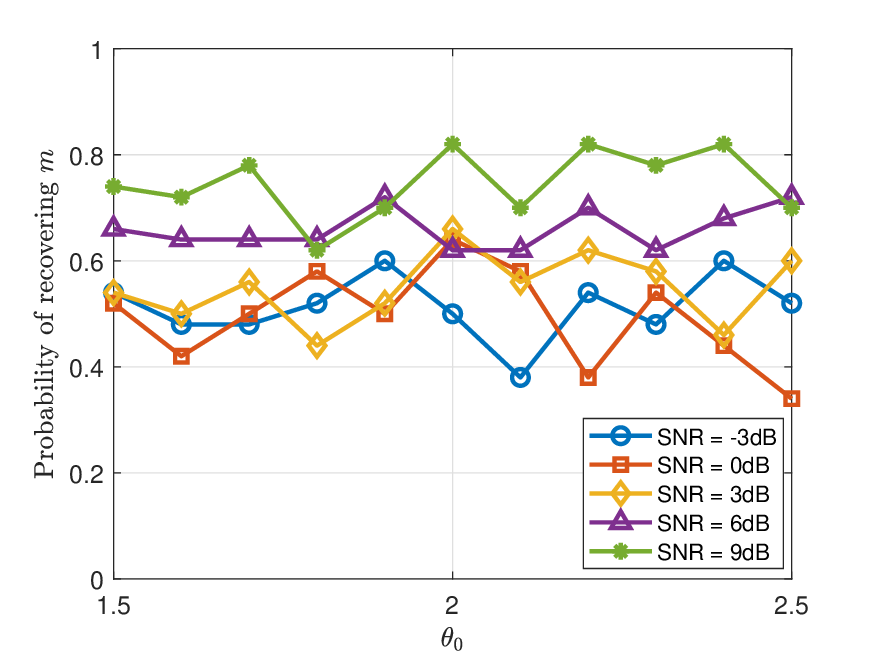}
			\caption{F-S ANM}
			\label{subfig:sims_Gfilt_SNR5}
		\end{subfigure}

		\caption{Panels (a)--(c): Probabilities of successfully recovering the number $m = 3$ of cisoids with our approach, the standard ANM, and the F-S ANM, respectively. Notice that the markers represent computed values and lines have no meaning.}
		\label{fig_ANMs_recov_compare}
	\end{figure}

\end{example}

\begin{example}\label{ex_compare_subspace}
	Now we compare our approach with two standard subspace methods, namely MUSIC and ESPRIT for frequency estimation. It is well known that subspace methods assume that the number $m$ of cisoids is known. In practice, however, such a parameter must be estimated by other methods. Here we use standard information criteria AIC and BIC for the estimation of $m$, see e.g., \cite{wax1985detection, stoica2005spectral}, and feed the estimated $\hat{m}$ to subspace methods.
        In addition, we also compare our approach which estimates $m$ by the numerical rank $\hat{r}$ of the optimal $\hat\Sigma$ (see Example~\ref{ex_separa_freqs}),
        with AIC and BIC  in recovering the correct number of cisoids.
        
	In this example, the signal component always contains seven cisoids ($m =7$) whose frequencies $\{\theta_k\}_{k=1}^7$ are randomly selected from the interval $[1.75, 2.25]$ meeting the condition that any two frequencies are separated at a distance no less than $\eta\Delta_\text{FFT}$ with a constant $\eta>0$ which we call a ``separation factor''. In fact, the smaller the factor $\eta$ is, the more challenging  the task of frequency estimation will be. The complex amplitudes are $a_k = e^{i\varphi_k}$ ($k =1, \dots, 7$) with phases $\varphi_k$ uniformly sampled from $[0, 2\pi]$. The definition of the $\SNR$ here is the same as that in Example~\ref{ex_compare_FS}.

        We resize the G-filter to have $n=30$ which selects the same frequency band $\Ical_1=[1.75, 2.25]$, with the same repeated pole $p=0.58e^{i2}$.
        We set the signal length of $y$ to be $L = 138$.	
        With the tolerance level $\varepsilon = 10^{-3}$, the number of filtered samples to be truncated is $L_{\srm} = 137$ which gives again $L-L_{\srm} = 1$. We fix the separation factor $\eta = 0.8$ and run 
        a Monte Carlo simulation for each choice of the $\SNR$ from the candidate set $\{2,3,\dots,8\}$ (dB). In Fig.~\ref{fig_AIC_BIC_recov_pros}, we display how the probabilities of recovering $m$ vary with the increasing $\SNR$ for AIC, BIC, and our approach. These plots demonstrate that our approach is uniformly better than AIC and BIC in detecting the number of cisoids $m=7$ over the selected range of the SNR.

		Next we study the accuracy of frequency estimation of our approach and the subspace methods MUSIC and ESPRIT used in conjunction with the AIC for estimating the number $m$ of cisoids. The use of AIC is suggested by Fig.~\ref{fig_AIC_BIC_recov_pros} where AIC performs better than BIC in $5$ of the $7$ instances of the SNR. The accuracy of frequency estimation is measured by the root mean square error ($\RMSE$)
		\begin{equation}\label{rmse_def}
			\text{$\RMSE$} = \sqrt{\frac{1}{N_{\text{sim}}} \sum_{j=1}^{N_{\text{sim}}} \frac{1}{m} \left\| \hat{\thetab}_j-\thetab_j \right\|_2^2}
		\end{equation}
		where \( N_{\text{sim}} \) is the number of trials that successfully recover the number of cisoids, \( \hat{\thetab}_j \) is the estimated frequency vector in the $j$-th trial, and \( \thetab_j \) is the corresponding true frequency vector.
		
		The first set of Monte Carlo simulations are done 
        for each value of the $\SNR$ from the candidate set $\{2,3,\dots,8\}$ (dB)
        with a fixed separation factor $\eta = 0.8$. Then the $\RMSE$ is calculated for different methods and the result is presented in Fig.~\ref{rmse_AIC_subspace}. It is shown that our approach uniformly outperforms ESPRIT and MUSIC in terms of the RMSE over the range of the SNR.
		The second set of Monte Carlo simulations are done for each choice of the separation factor $\eta$  from the arithmetic progression $\{0.5, 0.625, \dots, 1.5\}$ with a fixed $\SNR = 2$ dB.
        Then we compute the $\RMSE$ for each method and present the result in Fig.~\ref{fig_subfigure_m7}. 
        One can observe that the $\RMSE$ of our approach is the lowest when $0.5 \leq \eta \leq 1.375$ in comparison with the two subspace methods.

	\begin{figure}
		\centering
		\includegraphics[width=0.35\textwidth]{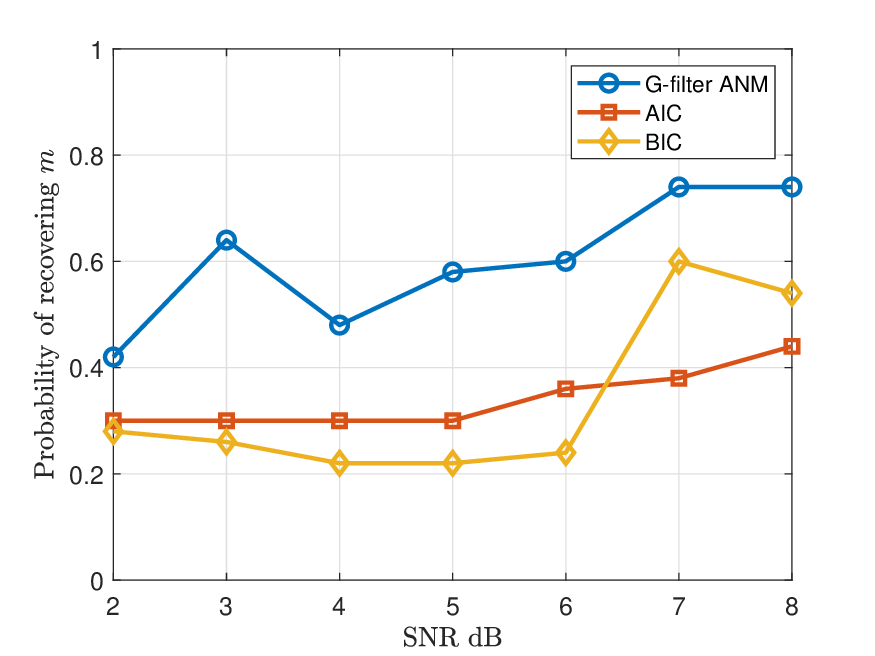}
		\caption{The ``probability'' of successfully recovering the number $m=7$ of cisoids versus the $\SNR$ while $\eta = 0.8$ is fixed. Such a probability is defined by the number of successful instances in a Monte Carlo simulation divided by $50$.  Notice that the markers represent computed values and lines have no meaning.}
		\label{fig_AIC_BIC_recov_pros}
	\end{figure}

	\begin{figure}
		\centering
		\includegraphics[width=0.35\textwidth]{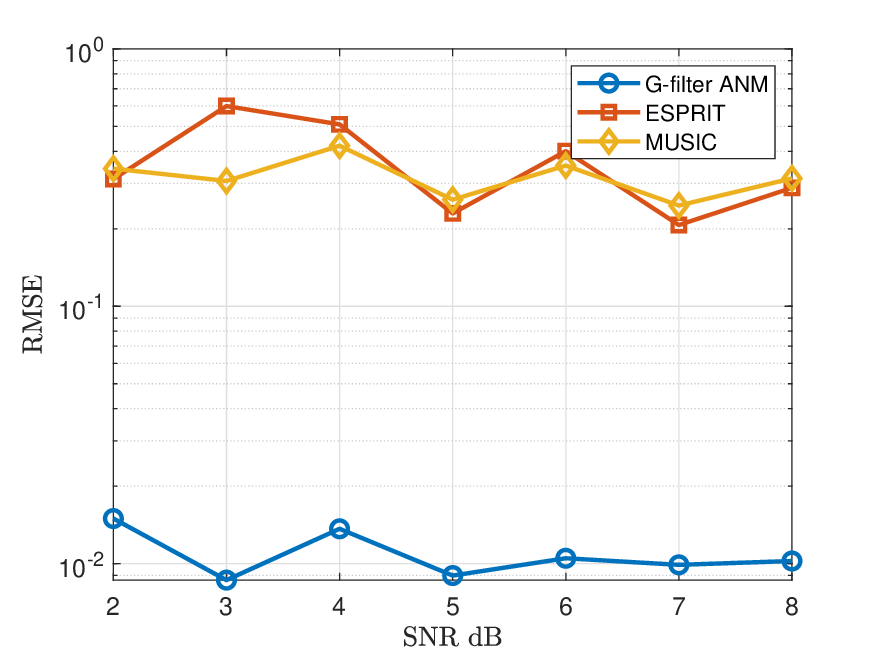}
		\caption{$\RMSE$ of frequency estimation versus the $\SNR$ while $\eta = 0.8$ is fixed. Notice that the markers represent computed values and lines have no meaning.}
		\label{rmse_AIC_subspace}
	\end{figure}

        	\begin{figure}
		\centering
		\includegraphics[width=0.35\textwidth]{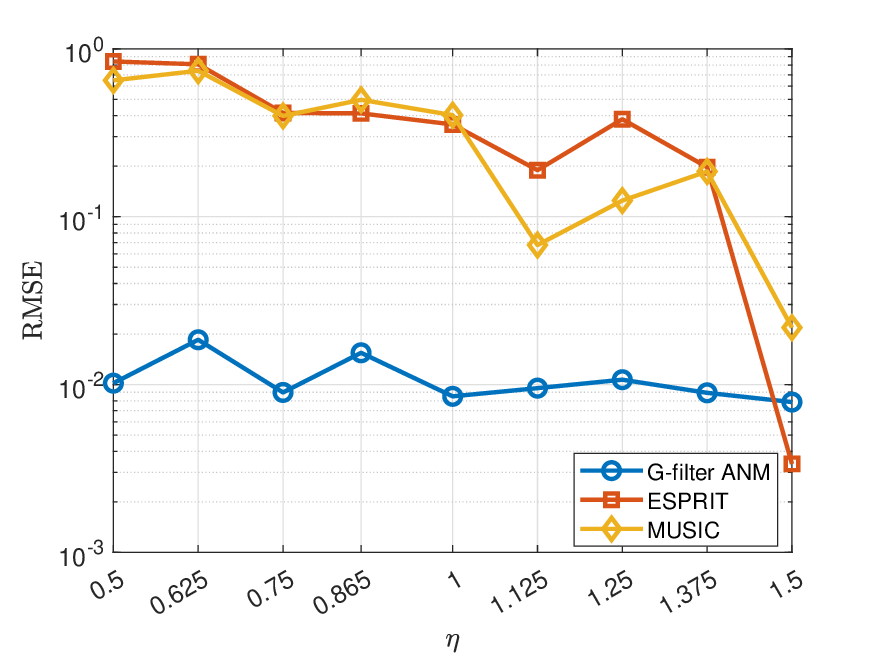}
		\caption{
        $\RMSE$ of frequency estimation versus the separation factor $\eta$ while $\SNR=2$ dB is fixed. Notice that the markers represent computed values and lines have no meaning.}
		\label{fig_subfigure_m7}
	\end{figure}
\end{example}

\section{Concluding remarks}\label{sec:conclus}

This paper describes an atomic norm minimization (ANM) framework for frequency estimation which generalizes the standard ANM theory by incorporating Georgiou's filter banks. The structure of state covariance matrices is exploited via a Carath\'{e}odory--Fej\'{e}r-type decomposition which is useful for frequency extraction. The optimization problem, which admits a semidefinite programming formulation, is investigated via dual analysis in the noiseless case, and the noisy case can be handled in a similar fashion using the atomic norm as regularization.
Results of numerical simulations show that the G-filter version of the ANM approach works 
{better than some alternative methods for frequency estimation}
provided that the SNR is not too low, and the true frequencies
are close to the band selected by the G-filter.

A number of open questions remain for future research: 
\begin{enumerate}
	\item {A proof of the existence of the dual certificate $Q(\theta)$ which satisfies \eqref{cond_interpola} and \eqref{cond_inequal}, see Remark~\ref{rem_exist_dual_certif}}.
	\item A tight estimate of the quantity $\E\|\tilde\wb\|_{\Acal}^*$ which can be used as the regularization parameter $\lambda$.
	\item A theoretical characterization of the resolution limit of the G-filter version of the ANM approach.
	\item Optimal design of the G-filter, i.e., choice of the parameter pair $(A,\bb)$ in the presence of prior knowledge of underlying frequencies.
\end{enumerate}

\appendix

\section{Auxiliary lemma}

Before stating the lemma, let us first examine in more detail the numerators of the vector-valued rational function $G(z)$ in \eqref{trans_func_filter_bank}. Using the cofactor formula for matrix inverse, we have
\begin{equation}\label{G_numerat_denomi}
\begin{aligned}
G(z) & = z(zI-A)^{-1} \bb \\
& = \frac{z\adj(zI-A)\bb}{\chi_A(z)}=:\frac{\nub(z)}{\chi_A(z)},
\end{aligned}
\end{equation}
where, $\chi_A(z):=\det(zI-A)$ is the (monic) characteristic polynomial of $A$, $\adj$ represents the adjugate matrix, and $\nub(z)=z\adj(zI-A)\bb$ is the vector-{valued} polynomial of degree at most $n$ with a constant term equal to zero. More precisely, we can write
\begin{equation}\label{G_numerator}
\nub(z) = \sum_{k=1}^{n} \nub_k z^k = N\psib(z)
\end{equation}
in which
\begin{itemize}
	\item $\nub_k\in\Cbb^n$ is a vector of polynomial coefficients,
	\item $N:=\bmat \nub_1 & \cdots & \nub_n \emat\in\Cbb^{n\times n}$,
	\item and $\psib(z) := \bmat z & \cdots & z^n\emat^\top$.
\end{itemize}
It is not difficult to infer from \citet[Proposition 1]{georgiou2000signal} that the square matrix $N$ is nonsingular and can be understood as a \emph{change-of-basis} matrix.

\begin{lemma}\label{lem_mat_repres_rat_func}
	Let $R\in\Hcal_n$ be such that $G^*(e^{i\theta}) R G(e^{i\theta})$ is a nonnegative rational function on $\Ical$ with $m$ distinct zeros $\theta_k$, $k=1, \dots, m$.
%	each having multiplicity one. 
	Then there exist $Y\in\ker\Gamma^*$ and a positive semidefinite matrix $H$ with rank $n-m$ such that
%	\begin{equation}
		$R+Y=H$.
%	\end{equation}	
	Moreover, $H$ has the maximal possible rank here.
\end{lemma}

\begin{pf}
	By \eqref{G_numerat_denomi} and \eqref{G_numerator}, the premise of the lemma implies that the symmetric polynomial
	\begin{equation}
		p(e^{i\theta}) := \nub^*(e^{i\theta}) R \nub(e^{i\theta}) = \psib^*(e^{i\theta}) N^* R N\psib(e^{i\theta})
	\end{equation}
	is nonnegative over $\Ical$, and $\theta_k\in\Theta$ are the zeros and there are precisely $m$ of them.
	By \citet[Lemma~B.1]{tang2013compressed}, there exists a positive semidefinite matrix $\tilde{H}$ of rank $n-m$ such that
	\begin{equation}
%		\begin{aligned}
			p(e^{i\theta}) = \psib^*(e^{i\theta}) \tilde{H} \psib(e^{i\theta}) = \nub^*(e^{i\theta}) H \nub(e^{i\theta}),
%		\end{aligned}
	\end{equation}
    where $H:=N^{-*}\tilde{H}N^{-1}$ is also positive semidefinite with rank $n-m$. Now the rational function
    \begin{equation}
    	\frac{p(e^{i\theta})}{|\chi_A(e^{i\theta})|^2} = G^*(e^{i\theta}) R G(e^{i\theta}) = G^*(e^{i\theta}) H G(e^{i\theta}),
    \end{equation}
    which implies that $Y:=H-R\in\ker\Gamma^*$, namely $G^*(e^{i\theta}) Y G(e^{i\theta})\equiv 0$. The maximality of the rank follows from similar reasoning to that in the proof of Lemma B.1 in \cite{tang2013compressed}, and this completes the proof.
	\pfend
\end{pf}

\bibliographystyle{model4-names}
\bibliography{references}           % and a bib file to produce the 
                                 % bibliography (preferred). The
                                 % correct style is generated by
                                 % Elsevier at the time of printing.

\end{document}